\documentclass[10pt,onecolumn,journal]{IEEEtran}
\usepackage{mathtools}
\usepackage{amssymb}
\usepackage{multirow}
\usepackage{stmaryrd}
\makeatletter
\input{Ustmry.fd}
\DeclareFontShape{U}{stmry}{b}{n}{<->ssub*stmry/m/n}{}
\makeatother
\usepackage{dsfont}
\usepackage{yfonts}
\usepackage{mathabx}
\usepackage{caption}
\usepackage{subcaption}
\usepackage{float}
\usepackage{stfloats}
\usepackage{graphicx}
\usepackage{tcolorbox}

\usepackage{tabularx}
\usepackage{booktabs}

\usepackage[numbers,sort&compress]{natbib}
\usepackage{multicol}
\usepackage{algorithm}
\usepackage{algpseudocode}
\usepackage{pifont}
\usepackage{varwidth}

\usepackage{afterpage}
\usepackage{balance}
\usepackage{mathbbol}

\def\mindex#1{\index{#1}}

\def\sq{\hbox{\rlap{$\sqcap$}$\sqcup$}}
\def\qed{\ifmmode\sq\else{\unskip\nobreak\hfil
\penalty50\hskip1em\null\nobreak\hfil\sq
\parfillskip=0pt\finalhyphendemerits=0\endgraf}\fi\medskip}

\long\def\defbox#1{\framebox[.9\hsize][c]{\parbox{.85\hsize}{%
\parindent=0pt
\baselineskip=12pt plus .1pt      
\parskip=6pt plus 1.5pt minus 1pt 
 #1}}}

\long\def\beginbox#1\endbox{\subsection*{}%
\hbox{\hspace{.05\hsize}\defbox{\medskip#1\bigskip}}%
\subsection*{}}

\def\endbox{}

\newsavebox{\junk}
\savebox{\junk}[1.6mm]{\hbox{$|\!|\!|$}}

\def\bfmath#1{{\mathchoice{\mbox{\boldmath$#1$}}%
{\mbox{\boldmath$#1$}}%
{\mbox{\boldmath$\scriptstyle#1$}}%
{\mbox{\boldmath$\scriptscriptstyle#1$}}}}

\def\bfmY{\bfmath{Y}}

\def\bfmhhaY{\bfmath{\hhaY}} 
\def\bfmhhaY{\hbox to 0pt{$\widehat{\bfmY}$\hss}\widehat{\phantom{\raise 1.25pt\hbox{$\bfmY$}}}}

\def\til={{\widetilde =}}

 \def\FRAC#1#2#3{\genfrac{}{}{}{#1}{#2}{#3}}

\def\ddtp{{\mathchoice{\FRAC{1}{d^{\hbox to 2pt{\rm\tiny +\hss}}}{dt}}%
{\FRAC{1}{d^{\hbox to 2pt{\rm\tiny +\hss}}}{dt}}%
{\FRAC{3}{d^{\hbox to 2pt{\rm\tiny +\hss}}}{dt}}%
{\FRAC{3}{d^{\hbox to 2pt{\rm\tiny +\hss}}}{dt}}}}

\def\average#1,#2,{{1\over #2} \sum_{#1}^{#2}}

\def\eye(#1){{\bf(#1)}\quad}

\def\eq#1/{(\ref{e:#1})}

\newcommand{\beqn}[1]{\notes{#1}%
\begin{eqnarray} \elabel{#1}}

\newcommand{\eeqn}{\end{eqnarray} }

\newcommand{\beq}[1]{\notes{#1}%
\begin{equation}\elabel{#1}}

\newcommand{\eeq}{\end{equation}}

\def\bdes{\begin{description}}
\def\edes{\end{description}}

\newcounter{rmnum}

\newcounter{anum}

{\end{list}}

\def\ass(#1:#2){(#1\ref{#1:#2})}

\def\ritem#1{
\item[{\sf \ass(\current_model:#1)}]
}

\newenvironment{recall-ass}[1]{%
\begin{description}
\def\current_model{#1}}{
\end{description}
}

\long\def\comment#1{}

\newfont{\bb}{msbm10 scaled 1100}

\newcommand{\nv}{{\bf n}}

\renewcommand{\arg}{{\hbox{arg}}}

\newcommand{\snrul}{{\sf snr}_{\rm ul}}

\newcommand{\betafb}{\beta_{\rm fb}}

\usepackage{tikz, pgfplots,graphicx,xcolor}
\usetikzlibrary{plotmarks,spy,backgrounds}
\pgfplotsset{compat=newest}

\allowdisplaybreaks

\title{Prior-Aided Masked Vector Quantization CSI Feedback for FDD Massive MIMO Systems}

\author{Yi Song$^1$, Tianyu Yang$^1$, Kangda Zhi$^1$, Shuangyang Li$^1$,
  and Giuseppe Caire$^1$%
  \thanks{$^1$Communications and Information Theory Group (CommIT), Technische Universit\"{a}t Berlin, 10587 Berlin, Germany (e-mail: \{yi.song, tianyu.yang, k.zhi, shuangyang.li, caire\}@tu-berlin.de).}}

\begin{document}
\maketitle

\begin{abstract}
  Downlink channel state information (CSI) feedback is a key bottleneck in
  frequency-division duplex (FDD) massive MIMO systems, as the user equipment
  (UE) must convey its estimated channel to the base station (BS) over a limited
  uplink (UL) budget. To improve CSI reconstruction accuracy under tight
  feedback constraints, we propose prior-aided masked vector quantization (PM-VQ),
  a learning-based separate source--channel coding (SSCC) feedback scheme
  conditioned on the average angle--delay power map---a compact
  representation of the channel second-order statistics available at both
  the UE and the BS. In PM-VQ, a prior-aided encoder maps the CSI to latent
  tokens, and a spatially-adaptive masking module (SAMM) scores and selects
  the most informative tokens within the feedback budget. The selected tokens
  are vector-quantized and fed back together with their positions, while an
  adaptive de-masking module (ADM) completes the latent representation at
  the BS before prior-conditioned decoding. To support variable-rate
  compression, a single model is trained over a range of selected-token
  counts, enabling operation across multiple feedback dimensions without
  retraining. We evaluate PM-VQ against three representative baselines on a
  Sionna-generated 3GPP TR~38.901 UMa dataset, focusing on the most
  challenging diffuse regime where channel energy is spread across many
  angle--delay coefficients. Simulation results show that PM-VQ achieves the
  lowest NMSE across all tested SNR levels and feedback dimensions in this
  regime. Moreover, the angle--delay power-map prior remains beneficial even
  when estimated from only a few channel realizations.
\end{abstract}

\begin{keywords}
  CSI feedback, FDD massive MIMO, vector quantization, side information,
  rate--distortion theory, masking, de-masking, deep learning.
\end{keywords}

\section{Introduction}
\label{sec:intro}
Accurate downlink (DL) channel state information (CSI) is the cornerstone of
next-generation wireless systems, since it directly determines the quality of
DL precoding and the achievable spectral efficiency. Acquiring such CSI with
low complexity, low latency, and limited overhead becomes increasingly critical
as systems scale toward more users, larger antenna arrays, and wider bandwidths.
How CSI is acquired depends on the duplexing mode. In time-division duplex
(TDD) systems, channel reciprocity allows the base station (BS) to infer the DL
channel directly from uplink (UL) pilots through an open loop, at the cost of
tighter transceiver synchronization and hardware calibration. Frequency-division
duplex (FDD) systems avoid this calibration burden but lack channel reciprocity,
which makes DL CSI acquisition intrinsically harder. In this case, a closed
feedback loop is required: the BS transmits common pilots, each user equipment
(UE) estimates the DL channel and feeds a compressed description back over the
UL, and the BS reconstructs the CSI from this feedback. This loop exposes a
fundamental tradeoff between pilot/feedback overhead and CSI reconstruction
accuracy. Designing feedback schemes that can accurately reconstruct
high-dimensional DL CSI under a limited UL feedback budget is the central
problem addressed in this paper.

\subsection{Uplink Feedback Strategies}

The CSI feedback sent from the UE to the BS over the UL can be designed in two
main ways, following the classical source--channel coding dichotomy
\cite{shannon1948mathematical}. In separate source--channel coding (SSCC), the
UE first compresses the DL CSI into a bitstream using a source encoder and then
protects these bits against UL noise using a channel code. The BS decodes the
received bits and reconstructs the CSI. In joint source--channel coding (JSCC),
the UE instead maps the CSI estimate directly onto UL channel symbols, and the
BS reconstructs the CSI from the noisy received symbols.
The two strategies are attractive for complementary reasons. SSCC is modular
and well understood: in the infinite-blocklength limit, a
rate--distortion-optimal source code followed by a capacity-achieving channel
code is optimal according to Shannon's source--channel separation theorem. 
{In practical finite-blocklength systems, however, this ideal behavior may
not be fully realized. For a given modulation and coding configuration, the
decoding-error probability can increase rapidly when the instantaneous UL
channel quality falls below its supported operating region, leading to the
familiar \emph{cliff effect}.}  
Moreover, the bit-level source-coding, channel-coding, and
decoding pipeline introduces additional latency. JSCC gives up this
modularity, but by mapping source representations directly to channel inputs, it
avoids an explicit bit interface and often degrades more gracefully with the UL
SNR in practical noisy-feedback regimes. Its main difficulty lies in code design:
information-theoretically optimal source--channel mappings exist in principle,
but are generally not available in closed form, and simple linear mappings can
be far from optimal in feedback-limited regimes.

Given these complementary tradeoffs, we adopt the SSCC framework for two reasons.
First, it is directly deployable: CSI reporting in FDD standards such as 5G NR is
digital, with the UE conveying bits over standardized, channel-coded uplink
control channels. An SSCC bitstream is compatible with this pipeline, whereas
JSCC transmits CSI as analog channel symbols that current digital feedback links
do not support. Second, and more fundamentally, the SSCC framework lets us treat source coding
and channel coding separately. By combining a rate--distortion-optimal source
code with a state-of-the-art, near-capacity channel code, the CSI feedback
problem reduces to a source-coding problem, so the design can concentrate on the
source encoder. This encoder can then be built around well-understood
rate--distortion principles.

{In a practical digital UL such as 5G NR, the rate at which information bits
can be reliably delivered to the BS over a given resource allocation depends
on the UL SNR. For a given UL SNR, this reliably supportable rate depends on
the employed modulation and coding scheme (MCS), the decoder implementation,
and the target block error rate. Through link adaptation, a higher UL SNR
generally permits a higher-order modulation and a higher code rate while
maintaining the prescribed reliability, whereas a lower UL SNR requires a more
robust configuration and therefore supports a lower information rate. Denoting
this reliably supportable rate by $R_{\rm ul}^{\rm ach}(\snrul)$, measured in
information bits per complex channel use, $\betafb$ complex UL feedback channel
uses provide a reliable feedback-bit budget satisfying
\begin{align}
    B \leq \betafb R_{\rm ul}^{\rm ach}(\snrul).
\end{align}}
{To study the CSI compression problem independently of a particular modulation,
channel code, and decoder implementation, we adopt an idealized
capacity-achieving channel-coding model. For the considered complex AWGN UL
feedback channel, the asymptotic upper limit on the reliably achievable rate is
\begin{align}
    C_{\rm ul}(\snrul)
    =
    \log_2\!\left(1+\snrul\right)
\end{align}
bits per complex channel use. With sufficiently long blocklengths and
capacity-approaching channel codes, the reliably supportable UL rate can
approach $C_{\rm ul}(\snrul)$. We therefore use the optimistic feedback budget
\begin{align}
    \label{eq:budget}
    B
    \leq
    \betafb C_{\rm ul}(\snrul).
\end{align}
Thus, for a fixed number of UL feedback channel uses $\betafb$, a higher UL SNR
increases $C_{\rm ul}(\snrul)$ and consequently increases the number of CSI
feedback bits that can be reliably delivered to the BS. This abstraction
reduces the noisy UL link to an optimistic reliable bit budget of
$\betafb C_{\rm ul}(\snrul)$ bits, allowing the comparison to focus on the CSI
source coding problem rather than on a particular UL channel-coding
implementation.}

\subsection{Related Work}

\paragraph{Theoretical guidelines}

The CSI feedback problem can be modeled as a remote source-coding problem,
where the UE must encode the DL CSI from the noisy DL channel observation under
a rate constraint relating to the UL feedback channel
capacity.
To study the information-theoretic limit, Khalilsarai et al.\
\cite{Khalilsarai2022channel,Khalilsarai2023FDD} derived, for a Gaussian
channel model, a lower bound on the CSI normalized mean-squared error (NMSE)
under a given rate constraint using remote rate--distortion theory.
The corresponding high-SNR scaling law characterizes how the error scales with
the DL pilot dimension, the UL feedback dimension, and the channel rank. This
result provides two useful design guidelines for CSI feedback. First, linear
JSCC achieves the optimal high-SNR scaling in all regimes except one: the regime
in which the DL pilot dimension exceeds the channel rank, while the UL feedback
dimension remains smaller than the rank. In this feedback-limited regime,
nonlinear encoding can in principle improve the scaling behavior. Second, for a
Gaussian source, {rate allocation of the
rate--distortion-optimal solution follows} reverse water-filling over the
eigenmodes of the source covariance, allocating more rate to high-variance
modes and no rate to modes below the water level. This suggests that an
efficient feedback encoder should exploit channel second-order statistics and
allocate the available feedback budget non-uniformly across channel
components.
Motivated by this observation, Song et al.\
\cite{song2025downlink,song2024joint} exploited the low-rank eigenspace of
spatially correlated channels \cite{jiang2015achievable} and proposed a
low-complexity JSCC scheme that encodes the dominant Karhunen--Lo\`eve
coefficients with optimized power allocation.
{This demonstrates that exploiting the channel covariance
structure can substantially simplify CSI feedback and guide the allocation of
limited feedback resources.}

{{However, an important gap remains between these asymptotic
information-theoretic characterizations and practical CSI feedback.
Rate--distortion achievability is inherently asymptotic: when each CSI matrix is
viewed as one realization of the source, the rate--distortion function $R(D)$ is
approached by jointly encoding blocks of source realizations, with the
asymptotic limit attained as the source-coding blocklength tends to infinity.
CSI feedback, in contrast, is naturally a \emph{one-shot} source-coding
problem: within each channel coherence interval, the UE must compress the
current CSI realization and convey it to the BS under a stringent latency
budget. The UE therefore cannot, in general, accumulate an arbitrarily long
sequence of CSI realizations and jointly encode them using an asymptotically
optimal rate--distortion code. Consequently, although
remote rate--distortion theory provides a fundamental performance benchmark
and reveals structural principles, such as reverse water-filling in the
Gaussian case, it does not directly specify how to realize these principles
with a practical finite-rate encoder operating on a single CSI realization.}}
{For the reasons explained before, we pursue these theoretical
guidelines within an SSCC framework.}
\emph{{The open question we address is therefore how to design
a trainable, variable-rate source encoder that operates effectively in
this one-shot regime, exploits channel second-order statistics, and allocates
the available feedback budget selectively across channel components on
realistic channel data.}}

\paragraph{Learning-based feedback}
The stringent latency and complexity requirements of CSI feedback in multi-user massive MIMO systems have motivated a broad class of learning-based feedback schemes. Learning-based JSCC maps the CSI representation directly to channel inputs, and the BS reconstructs the CSI from the noisy received symbols.
Early work adopted CNN-based encoder--decoder architectures for CSI feedback
\cite{mashhadi2020cnn}. Subsequent studies improved robustness by introducing
SNR-adaptive transmission strategies \cite{xu2022deep} and feedback-aided
variants that exploit channel-output feedback \cite{kurka2020deepJSCCf}. A
comprehensive overview of recent JSCC developments can be found in
\cite{gunduz2025joint}.
In parallel, a large body of work has studied learning-based SSCC CSI
compression. A representative example is CsiNet \cite{Wen2017DeepLF}, which
employs a CNN autoencoder to compress and reconstruct angle--delay CSI. Later
extensions improved the reconstruction quality by enlarging the receptive field
\cite{guo2020convolutional}, introducing multi-resolution feature extraction
\cite{lu2020multi}, and incorporating generative priors
\cite{tolba2020massive}. An overview of learning-based SSCC feedback schemes is
provided in \cite{guo2022overview}. More recently, attention-based
architectures, such as TransNet \cite{cui2022transnet}, have been introduced to
capture long-range dependencies in CSI and are therefore adopted in this paper
as the backbone of the neural network baselines.

\paragraph{Limitations}
Despite these advances, existing learning-based CSI feedback schemes still have
several limitations that motivate our design. First, since latent
representations are often learned purely from data, different latent
components may carry different amounts of information about the CSI. As a
result, the feedback budget should concentrate on the most informative components. However, most existing schemes do not explicitly identify or prioritize the most informative components under the feedback budget.
Second, although rate--distortion theory highlights the importance of channel second-order statistics, most learning-based schemes do not explicitly exploit it in the compression and reconstruction process. Zhuang et al.\ \cite{zhuangcovnet2025} exploited the channel second-order statistics but only at the decoder for reconstruction.
Together, these two limitations suggest that the encoder should not only learn a compact CSI representation, but also \emph{identify and prioritize statistically important latent components}.
Third, most learning-based CSI feedback schemes support only a fixed compression ratio, requiring a separate model for each feedback budget. Existing variable-rate methods typically construct nested latent representations and transmit the first $k$ features according to the available resources \cite{lee2024learning, ankireddy2026residual, valentina2023user}. However, their feature ordering is generally fixed across channel realizations.
Fourth, most existing evaluations rely on datasets generated from a single fixed channel profile, such as the standardized CDL-C or CDL-D models \cite{3gpp38901}, with limited variation in delay profile, angular spread, array geometry, and user layout, which limits conclusions about generalization when the deployment environment differs from the training distribution.

\subsection{Contributions}

Motivated by the above limitations, we propose a learning-based SSCC feedback
scheme, called prior-aided masked vector quantization (PM-VQ), for DL CSI
feedback. Inspired by the reverse water-filling structure of the rate--distortion-optimal solution, PM-VQ uses the
average angle--delay power map as prior side information to guide non-uniform
latent-token selection under a given feedback budget. It combines
vector-quantized latent compression with prior-aware token masking and
mask-based latent completion, so that the encoder transmits only the most
informative latent tokens and the BS reconstructs the full channel from the
received tokens, with the channel statistical prior injected at the encoder,
token-scoring module, and decoder. The main contributions are summarized as
follows.

\begin{itemize}

  \item \textbf{Multi-level prior conditioning.} We exploit the average
    angle--delay power map as prior side information and inject it at multiple
    semantic levels: gated fusion in the encoder, {SPADE}-style modulation for
    token scoring and decoding, and BS-side prior skip connections. This
    conditions both compression and reconstruction on the channel
    second-order statistics.

  \item \textbf{Selective latent-token transmission.} We introduce a
    spatially-adaptive masking module (SAMM) that scores latent tokens using both
    the instantaneous CSI and the long-term prior and retains only the most
    informative ones under the available feedback budget. At the BS, an adaptive
    de-masking module (ADM) reconstructs the masked tokens from the received
    quantized tokens.

  \item \textbf{Content-adaptive variable-rate feedback.} Because SAMM selects a
    variable number of tokens per channel realization, a single trained model
    operates across multiple feedback dimensions $\beta_{\rm fb}$ without retraining, and the selection
    adapts to each individual CSI sample.

  \item \textbf{Geometrically diverse dataset.} We generate a DL CSI dataset with
    the Sionna simulator under the 3GPP TR~38.901 UMa model, sampling diverse UE
    locations around a fixed BS to introduce geometric diversity beyond
    single-profile training.

  \item \textbf{Unified benchmarking.} We compare PM-VQ and all baselines under a common UL feedback dimension $\beta_{\rm fb}$ measured in channel uses. On a UMa dataset spanning channel-rank conditions from diffuse to sparse, we
focus the evaluation on the most challenging diffuse regime, where the channel
energy spreads across many angle--delay coefficients. Extensive simulations show
that PM-VQ achieves the best NMSE across SNR levels and feedback dimensions $\beta_{\rm fb}$ in this
regime, and its advantage over the baselines is especially pronounced. The
results also demonstrate that the prior provides substantial gains even when the
angle--delay power map is itself noisy.

\end{itemize}

The remainder of this paper is organized as follows.
Section~\ref{sec:system_model} introduces the FDD massive MIMO-OFDM system
model, the CSI representation, and the prior information. Section~\ref{sec:proposed_method} presents the proposed PM-VQ feedback
scheme. Section~\ref{sec:dataset} details the dataset generation procedure,
Section~\ref{sec:baselines} describes the
baseline methods, Section~\ref{sec:results} presents the simulation results, and
Section~\ref{sec:conclusion} concludes the paper.

\section{System Model}
\label{sec:system_model}

We consider an FDD massive MIMO-OFDM system, where a BS equipped with an $N_{\mathrm{BS}}$-element uniform linear array (ULA) serves multiple UEs, each also equipped with an $N_{\mathrm{UE}}$-element ULA. Since the UEs estimate and feed back their channels independently, we focus on a single generic UE and omit the user index throughout.
Let $\mathcal{F}$ denote the set of selected OFDM subcarriers, with
$N_{\mathrm{sc}}=|\mathcal{F}|$. At subcarrier $f\in\mathcal{F}$, let $\mathbf{H}_{f}\in\mathbb{C}^{N_{\mathrm{UE}}\times N_{\mathrm{BS}}}$ denote the DL channel.
To isolate the feedback problem from pilot design and channel estimation, we
assume the UE has acquired its DL channel $\{\mathbf{H}_{f}\}$ \emph{perfectly}
before feedback. Under this assumption, the remote source-coding formulation of
Section~\ref{sec:intro} reduces to direct compression of the known channel. Our
objective is a \emph{one-shot} feedback scheme that compresses the CSI at the UE
and reconstructs it at the BS under a limited UL feedback budget.

\subsection{Angle--Delay CSI Representation}
Rather than operating in the antenna--frequency domain, we represent the CSI in its angle--delay dual domain, where the channel sparsity becomes apparent due to the limited delay and angular spread of the propagation paths, which can be leveraged for channel compression.
For UE antenna $r$, let
$\mathbf{H}^{(r)}\in\mathbb{C}^{N_{\mathrm{sc}}\times N_{\mathrm{BS}}}$ collect
its antenna--frequency channel over the selected subcarriers and BS antennas.
Its angle--delay representation is the two-sided unitary DFT
\begin{equation}
  \mathbf{H}_{\rm {ad}}^{(r)}=\mathbf{F}_{N_{\mathrm{sc}}}^{H}\,\mathbf{H}^{(r)}\,\mathbf{F}_{N_{\mathrm{BS}}},
  \qquad r=1,\ldots,N_{\mathrm{UE}},
  \label{eq:angle_delay}
\end{equation}
where $\mathbf{F}_{N}$ is the unitary $N$-point DFT matrix. Left multiplication by
$\mathbf{F}_{N_{\mathrm{sc}}}^{H}$ maps frequencies to delays, and right
multiplication by $\mathbf{F}_{N_{\mathrm{BS}}}$ maps BS antennas to angles.
Stacking $\{\mathbf{H}_{\rm ad}^{(r)}\}_{r=1}^{N_{\mathrm{UE}}}$ gives the
angle--delay CSI tensor
$\mathbf{H}_{\rm ad}\in\mathbb{C}^{N_{\mathrm{sc}}\times N_{\mathrm{BS}}\times N_{\mathrm{UE}}}$.
Being unitary, this transform preserves channel energy while exposing the
channel's sparse delay--angular structure.

\subsection{Prior Information: Average Angle--Delay Power Map}
\label{sec:prior}

To compress and reconstruct the channel more effectively,
we exploit the channel second-order statistics
as prior information. Since estimating and storing
the full covariance is costly in massive MIMO systems, we
instead approximate it by the average power in the angle–delay domain. In sparse multipath channels, the DFT basis
approximately de-correlates the angle–delay coefficients, so
the per-coefficient power captures the dominant second-order
structure.

{The resulting average power map admits a natural interpretation in
terms of power spectral density (PSD). For a wide-sense stationary random
process, the PSD is the Fourier transform of its autocorrelation function and
describes how the average signal power is distributed over frequency.
Analogously, applying the DFT along the spatial and frequency dimensions
transforms the channel into discrete angle--delay representation. The quantity
$\mathbb{E}[|\mathbf{H}_{\rm ad}|^2]$ \footnote{$|\cdot|^{2}$ is applied elementwise} therefore describes how the average
channel power is distributed across the angle--delay bins and can be viewed as a
finite-grid angle--delay power spectrum. For a single realization,
$|\mathbf{H}_{\rm ad}(s)|^2$ gives the instantaneous power distribution,
whereas averaging over multiple realizations estimates the underlying
slowly varying power spectrum.}

{Specifically, we
estimate the \emph{average angle--delay power map} from $S$ independent small-scale realizations that share the same
large-scale statistics (e.g., the same UE location),}
\begin{equation}
  \mathbf{P}^{(S)}=\frac{1}{S}\sum_{s=1}^{S}\bigl|{\mathbf{H}_{\rm ad}(s)}\bigr|^{2}
  \;\in\;\mathbb{R}^{\,N_{\mathrm{sc}}\times N_{\mathrm{BS}}\times N_{\mathrm{UE}}},
  \label{eq:prior}
\end{equation}
{Thus, $\mathbf{P}^{(S)}$ is a sample-average estimate of the
underlying angle--delay power spectrum
$\mathbb{E}[|\mathbf{H}_{\rm ad}|^2]$. Such averaging is feasible because
small-scale fading typically varies faster than the geometry-induced
second-order statistics, so multiple channel realizations can exhibit
different instantaneous fading while sharing approximately the same
angle--delay power distribution.}

We assume that both the UE and the BS maintain a buffer of historical
angle--delay CSI: the UE accumulates its own channel estimates, while the BS
accumulates the CSI it has already reconstructed from earlier feedback rounds.
Each side then computes the power map directly from its local buffer via
\eqref{eq:prior}, so the prior becomes available at both ends without any
additional feedback payload.
Because the two
buffers are formed from correlated realizations of the same slowly varying
large-scale statistics, the resulting estimates are consistent, and the residual
mismatch between them is tolerated by the robustness of PM-VQ to prior-estimation
error demonstrated in Section~\ref{sec:results}.

\section{Proposed Prior-Aided Masked Vector Quantization CSI Feedback Scheme}
\label{sec:proposed_method}
\begin{figure*}[t]
  \centering
  \includegraphics[width=0.98\textwidth]{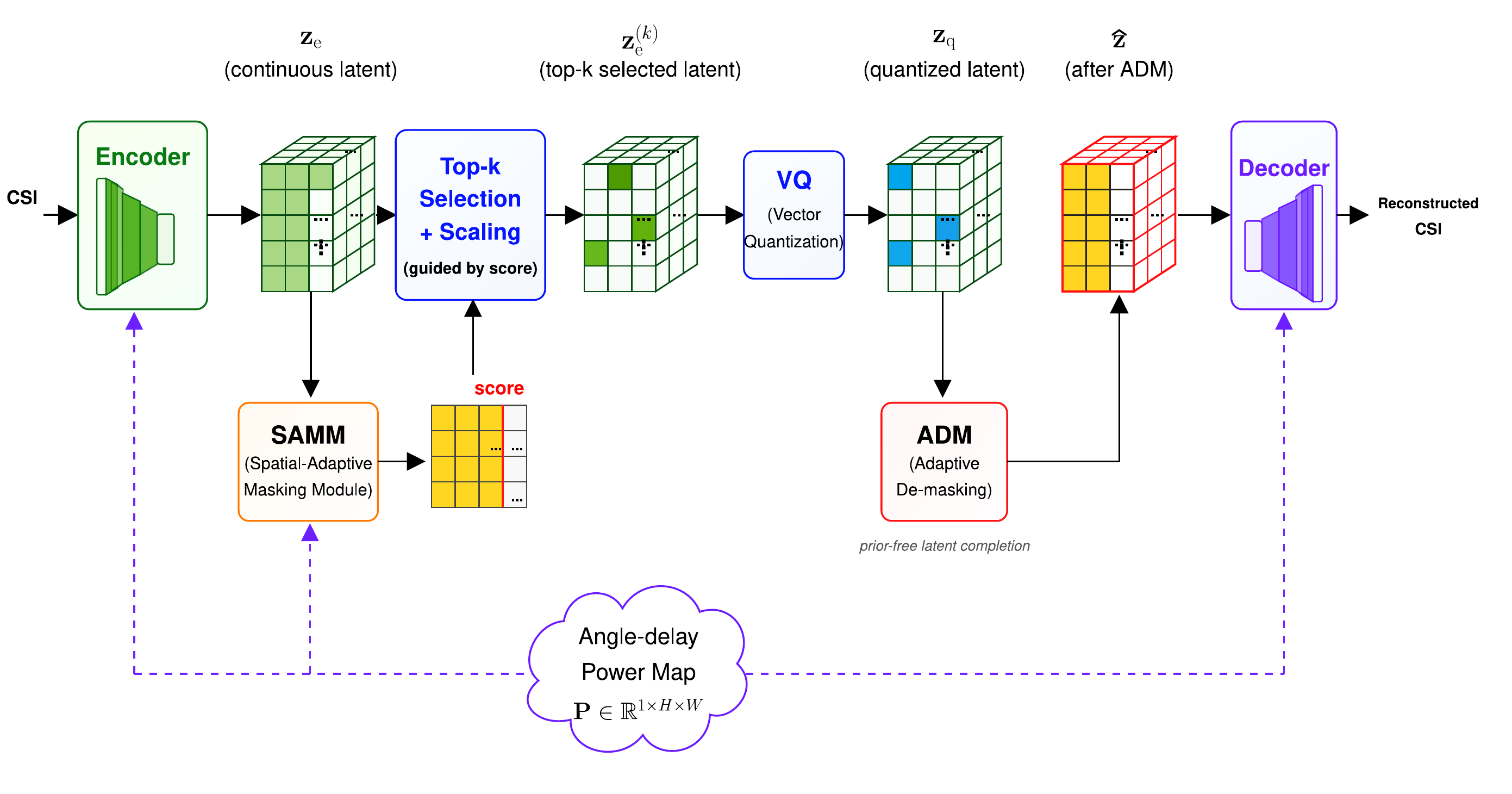}
  \caption{System diagram of the proposed PM-VQ framework.}
  \label{fig:system}
  \vspace{-5mm}
\end{figure*}

In this section, we present the proposed CSI feedback scheme, a learnable
SSCC-based design conditioned on prior side information. As shown in
Fig.~\ref{fig:system}, the scheme builds on the encoder--decoder framework with
vector quantization (VQ): the encoder at the UE compresses the DL CSI into a
latent representation, the latent vectors are quantized to indices of a learned
codebook and fed back, and the decoder at the BS reconstructs the DL CSI from
the received indices. This framework is standard in learning-based CSI feedback
\cite{guo2022overview}, but it suffers from three limitations:
\begin{itemize}
  \item The network must be retrained for every target compression ratio,
    since the latent dimension is fixed by the architecture.
  \item The compression mapping is purely data-driven: it exploits the
    redundancy of the high-dimensional CSI, but provides no measure of which
    latent dimensions are more informative and should be prioritized under a
    limited feedback budget.
  \item There is no principled mechanism for integrating side information
    into the compression and reconstruction designs.
\end{itemize}
The proposed prior-aided masked vector quantization (PM-VQ) feedback scheme
addresses these limitations one by one: it supports a range of feedback dimensions
with a single trained model by varying the number of transmitted latent tokens;
it scores the latent tokens so that only the most informative ones are fed
back; and it incorporates prior side information, i.e., the average angle--delay
power map introduced in Section~\ref{sec:prior}, which is assumed to be available at both the UE and the
BS, at multiple points in the encoder, SAMM, and decoder. The scheme consists of the following
components:
\begin{itemize}
  \item \textbf{Prior-aided encoder}: compresses the DL CSI into latent
    tokens, conditioned on the prior;
  \item \textbf{Spatially-adaptive masking module (SAMM)}: assigns an
    importance score to each latent token based on both the tokens and the
    prior power map;
  \item \textbf{Top-$k$ selection and scaling}: retains the $k$
    highest-scoring tokens and scales each retained token by its score;
  \item \textbf{Vector quantization}: quantizes the selected, scaled tokens;
    the codeword indices and the positions of the selected tokens are
    transmitted to the BS over the UL feedback channel;
  \item \textbf{Adaptive de-masking module (ADM)}: completes the latent
    representation at the BS from the received codewords and positions;
  \item \textbf{Prior-aided decoder}: reconstructs the DL CSI from the
    completed latent representation, again conditioned on the prior.
\end{itemize}

A further design principle of PM-VQ is that the prior is \emph{not} merely
concatenated with or added to the CSI at the input. Instead, it is injected at
multiple semantic levels:
\begin{itemize}
  \item[(i)] gated additive fusion at multiple encoder resolutions,
  \item[(ii)] SPADE-style modulation \cite{Park2019SemanticIS} for token scoring and decoding,
  \item[(iii)] a \emph{BS-side} prior encoder that synthesizes U-Net skip
    connections without violating the CSI-feedback constraint that UE-side
    features are unavailable at the BS.
\end{itemize}
These injection points correspond to the dashed arrows from the power map to
the encoder, SAMM, and the decoder in Fig.~\ref{fig:system}.

\subsection{Asymmetric Prior-Aided U-Net Backbone}
\label{sec:overview}

The encoder--decoder backbone is motivated by two properties of massive
MIMO-OFDM CSI in the angle--delay domain. First, the angle--delay channel
behaves as a structured image whose energy concentrates in a small number of
localized clusters. Accurate reconstruction therefore requires both global
context, which locates the dominant clusters, and local detail, which describes
the coefficients within each cluster. Second, the average angle--delay power
map $\mathbf{P}$ already indicates the likely locations of the dominant paths
and is available at both the UE and the BS.

The first property motivates a U-Net-like multi-scale architecture
\cite{ronneberger2015u}, in which the encoder progressively reduces the spatial
resolution so that the bottleneck latent captures the global channel structure,
while the decoder progressively restores the resolution to recover fine
angle--delay detail. The second property determines how the architecture
is modified from a standard symmetric U-Net. In a conventional U-Net,
intermediate encoder features are passed directly to the decoder through skip
connections, but in CSI feedback these features are computed at the UE, and
transmitting them to the BS would consume additional feedback resources. We
therefore adopt an \emph{asymmetric} U-Net, where the UE transmits no
intermediate features, and the BS decoder replaces the missing encoder skips
with multi-scale features extracted from the prior power map, conveying similar
structural information at zero feedback cost.

\subsection{Input Representation}
\label{sec:input}
We treat the complex angle--delay channel $\mathbf{H}_{\rm ad}$ as a two-channel
real-valued image $\mathbf{X}\in\mathbb{R}^{2\times H\times W}$, stacking its
real and imaginary parts, where the height $H$ indexes delays and the width
$W=N_{\mathrm{BS}}N_{\mathrm{UE}}$ indexes the flattened angle
pairs. The prior power map $\mathbf{P}^{(S)}$ is reshaped and normalized to the
same layout, giving $\mathbf{P}\in\mathbb{R}^{1\times H\times W}$.
The two inputs are complementary: $\mathbf{X}$ is the instantaneous realization to be
compressed, while $\mathbf{P}$ summarizes the slowly varying second-order
geometry of the channel.

\subsection{Prior-Aided Encoder with Gated Fusion}
\label{sec:encoder}
The encoder maps the CSI image and the prior to a compact latent feature map,
\begin{equation}
  \mathbf{z}_{\rm e}=E_{\theta}(\mathbf{X},\mathbf{P})\in\mathbb{R}^{C\times h\times w},
  \qquad K=hw,
  \label{eq:encode}
\end{equation}
where $C$ is the latent channel dimension and $K=hw$ the number of spatial
latent tokens, written $\mathbf{z}_{{\rm e},i}\in\mathbb{R}^{C}$ for position
$i$. Architecturally, $E_{\theta}$ is the contracting path of an asymmetric
U-Net: a $3\times3$ convolution, $n_{\mathrm{down}}$ residual downsampling
stages, and a $1\times1$ projection to dimension $C$, so that successive
downsampling aggregates increasingly global angle--delay context.

Rather than conditioning only on the input, we inject the prior at every encoder
resolution. At stage $\ell$, with CSI feature map $\mathbf{f}_{\ell}$, the prior
is resized to the same resolution, projected by a small network
$\rho_{\ell}(\cdot)$, and fused as
\begin{equation}
  \mathrm{Fuse}_{\ell}(\mathbf{f}_{\ell},\mathbf{P})
  =\mathbf{f}_{\ell}+\tanh(\alpha_{\ell})\,\rho_{\ell}(\mathbf{P})+\mathrm{GN}(\mathbf{f}_{\ell}),
  \label{eq:enc_gated_add}
\end{equation}
with a trainable scalar $\alpha_{\ell}$ and GroupNorm $\mathrm{GN}(\cdot)$. Each of
the three terms serves a distinct purpose. The identity term $\mathbf{f}_{\ell}$
passes the CSI features through unchanged, preserving their amplitude
information. The normalized term $\mathrm{GN}(\mathbf{f}_{\ell})$ adds a
scale-normalized version of the same features, which keeps activations at a
consistent magnitude across different channel realizations and network layers and
thus makes training more stable. The prior term $\rho_{\ell}(\mathbf{P})$ adds the
long-term angle--delay power structure, resized to match the current resolution. The bounded gate $\tanh(\alpha_{\ell})$ lets each stage learn
how strongly to rely on the prior, adding statistical guidance without
overwhelming the instantaneous CSI.

\subsection{Spatially-Adaptive Masking Module}
\label{sec:samm}
Once the latent is formed, the UE must decide which tokens to transmit under the
budget. The spatially-adaptive masking module (SAMM) is a lightweight network
that scores each latent token,
\begin{equation}
  \mathbf{s}=\mathrm{SAMM}(\mathbf{z}_{\rm e},\widetilde{\mathbf{P}})\in(0,1)^{K},
  \label{eq:samm_score}
\end{equation}
where $\widetilde{\mathbf{P}}$ is the prior downsampled to the latent resolution
and refined by a small CNN, and a higher score marks a token as more useful for
reconstruction. SAMM conditions on the prior through spatially-adaptive
normalization (SPADE) \cite{Park2019SemanticIS}. For a feature map $\mathbf{h}$,
\begin{equation}
  \mathrm{SPADE}(\mathbf{h},\widetilde{\mathbf{P}})
  =\bigl(1+\boldsymbol{\gamma}(\widetilde{\mathbf{P}})\bigr)\odot\mathrm{GN}(\mathbf{h})
  +\boldsymbol{\beta}(\widetilde{\mathbf{P}}),
  \label{eq:spade}
\end{equation}
with modulation maps $\boldsymbol{\gamma}(\cdot),\boldsymbol{\beta}(\cdot)$
produced from the prior by a small convolutional network. Unlike concatenation,
SPADE reinjects the prior after normalization and thus modulates the feature
statistics directly according to the dominant angle--delay clusters. Given the
scores, the UE keeps the $k$ highest-scoring tokens,
\begin{equation}
  \mathcal{I}=\operatorname*{top\text{-}k}(\mathbf{s}),
  \qquad m_i=\mathds{1}\{i\in\mathcal{I}\},
  \label{eq:topk}
\end{equation}
and discards the remaining $K-k$ latent tokens. SAMM therefore performs content-adaptive
selection: for each realization it spends the feedback budget on the latent
positions predicted to be most informative.

\subsection{Score-Scaled Vector Quantization}
\label{sec:vq}
The selected tokens are mapped to feedback bits by vector quantization. Before
quantizing, we normalize each token with a token-wise LayerNorm and then scale it
by its SAMM score,
\begin{equation}
  \widetilde{\mathbf{z}}_{{\rm e},i}=\mathrm{LN}(\mathbf{z}_{{\rm e},i})\,s_i,
  \qquad i\in\mathcal{I}.
  \label{eq:score_scale}
\end{equation}
The LayerNorm stabilizes the distribution seen by the quantizer. The score
scaling serves a different purpose. Since the top-$k$ selection that picks the
tokens is non-differentiable, multiplying by $s_i$ is what places the score on
the forward computation path, so that the reconstruction loss can back-propagate
into the SAMM scores and train the selection network. In effect, the scaling is
the bridge that makes the otherwise discrete token selection trainable
end-to-end.
Quantization uses a learned
codebook $\mathcal{C}=\{\mathbf{c}_{j}\}_{j=1}^{J}$ with
$\mathbf{c}_{j}\in\mathbb{R}^{C}$, assigning each token to its nearest codeword,
\begin{equation}
  q_i=\arg\min_{j}\bigl\|\widetilde{\mathbf{z}}_{{\rm e},i}-\mathbf{c}_{j}\bigr\|_{2}^{2},
  \qquad \mathbf{z}_{q_i}=\mathbf{c}_{q_i}.
  \label{eq:vq}
\end{equation}
Gradients pass through the non-differentiable assignment via the
straight-through estimator,
\begin{equation}
  \widehat{\mathbf{z}}_{q_i}=\widetilde{\mathbf{z}}_{{\rm e},i}
  +\mathrm{sg}\bigl(\mathbf{z}_{q_i}-\widetilde{\mathbf{z}}_{{\rm e},i}\bigr),
  \label{eq:vq_ste}
\end{equation}
so the forward pass uses the codeword while the backward pass updates the
score-scaled encoder token.

\subsection{Feedback Budget and Number of Kept Tokens}
\label{sec:budget}
The number of retained tokens follows directly from the feedback budget. The UE
should transmit both the codeword indices $\{q_i\}_{i\in\mathcal{I}}$ and the selected
positions $\mathcal{I}$. The positions cannot be inferred at the BS from the
prior alone, since the mask is computed from the UE-side latent. With a codebook
of size $J$, the indices cost $k\log_2 J$ bits, while describing the support of
$k$ tokens among $K$ positions costs $\log_2\binom{K}{k}\le K\,h_2(k/K)$ bits,
where $h_2(\cdot)$ is the binary entropy. We therefore model the source-coding
cost as
\begin{equation}
  B(K,k,J)=K\,h_2\!\left(\tfrac{k}{K}\right)+k\log_2 J,
  \label{eq:src_bits}
\end{equation}
which upper-bounds the exact positional cost, so the budget constraint is
always respected.
Given the reliable budget $\betafb C_{\mathrm{ul}}$ of
Section~\ref{sec:intro}, we retain the largest admissible number of tokens,
\begin{equation}
  k^{\star}=\max\bigl\{k:\,B(K,k,J)\le\betafb C_{\mathrm{ul}}\bigr\}.
  \label{eq:k_star}
\end{equation}
Since $\betafb$, $C_{\mathrm{ul}}$, $K$, and $J$ are all known at both the UE
and the BS, $k^{\star}$ is determined identically at both sides and does not
need to be signaled. The feedback resource thus sets $k^{\star}$ directly, and a
single trained model serves any budget by varying $k^{\star}$, with no
architectural change or retraining.

\subsection{Adaptive De-Masking Module}
\label{sec:adm}
When the BS receives the $k^{\star}$ selected and quantized tokens
$\mathbf{z}_{q}=\{\mathbf{z}_{q_i}\}_{i\in\mathcal{I}}$ and their grid positions
$\mathcal{I}$, the adaptive de-masking module (ADM) aims to reconstruct the
complete set of latent tokens
$\widehat{\mathbf{z}}\in\mathbb{R}^{C\times h\times w}$ required by the decoder.
To achieve this, ADM first restores the full $K$-token
grid, placing each received codeword at its own position and marking every
missing position with a single shared, learnable mask token,
\begin{equation}
  \widetilde{\mathbf{z}}_{0,i}=
  \begin{cases}
    \mathbf{z}_{q_i}, & i\in\mathcal{I},\\
    \mathbf{z}_{\mathrm{mask}}, & i\notin\mathcal{I},
  \end{cases}
  \label{eq:adm_sparse_grid}
\end{equation}
which turns the problem into one of completing the missing positions from the
received ones.
{Let $\mathbf{m}\in\{0,1\}^{K}$ be the binary indicator of the
selected positions, with $m_{i}=1$ if $i\in\mathcal{I}$ and $m_{i}=0$ otherwise,
so that $\mathbf{m}$ marks the received and missing entries of the grid. The
completion is then written as}
\begin{equation}
  \widehat{\mathbf{z}}=\mathrm{ADM}\bigl(\widetilde{\mathbf{z}}_0,{\mathbf{m}}\bigr).
  \label{eq:adm}
\end{equation}

Following masked generative models \cite{chang2022maskgit}, ADM splits this
completion into two parts, a fixed baseline plus a learned correction. The
baseline is a dimension-preserving projection of the reassembled grid,
\begin{equation}
  \mathbf{b}=\mathcal{P}_{0}(\widetilde{\mathbf{z}}_{0})
  \label{eq:adm_anchor}
\end{equation}
which gives the received codewords a direct path to the output, so their
information is preserved rather than reconstructed from scratch. On top of this
baseline, a transformer branch predicts the correction needed to fill in the
missing positions,
\begin{equation}
  \Delta\mathbf{z}=\mathcal{R}_{\psi}\!\left(\mathcal{T}_{\psi}\!\left(\mathbf{b},\{\mathbf{a}^{(\ell)}\}_{\ell=0}^{L-1}\right)\right),
  \label{eq:adm_residual}
\end{equation}
where $\mathcal{T}_{\psi}$ stacks $L$ masked self-attention blocks and
$\mathcal{R}_{\psi}$ is a convolutional head. The ADM completes the latent purely
from the received tokens and does not use the prior; the missing positions are
inferred from the reliable received tokens through the completion mechanism
described next.
A key difficulty is that the mask tokens are only placeholders: they hold no
channel information, yet they sit on the same grid as the received tokens. When
the network reconstructs a position, it does so by combining information from all
positions on the grid. If the placeholders are given the same weight as the
received tokens in this combination, their empty content leaks into the result
and corrupts the reconstruction. The network must therefore weigh the received
tokens far more heavily than the placeholders. We address this with a \emph{progressive
reliability mask}. For each layer $\ell$, we build a soft
reliability weight $\mathbf{a}^{(\ell)}$, that sets how much
layer $\ell$ should trust each position. Initially the weight strongly favors the received tokens,
\begin{equation}
  \mathbf{a}^{(0)}=\mathbf{m}+\epsilon(1-\mathbf{m}),
  \qquad 0<\epsilon\ll1,
  \label{eq:adm_progressive_mask_init}
\end{equation}
giving received positions weight $1$ and missing positions a near-zero weight
$\epsilon$. After each layer the weight is raised toward $1$ by taking a square
root,
\begin{equation}
  \mathbf{a}^{(\ell+1)}=\sqrt{\mathbf{a}^{(\ell)}}.
  \label{eq:adm_progressive_mask}
\end{equation}
Since repeatedly square-rooting any value in $(0,1]$ pushes it toward $1$, the
missing positions become gradually more trusted as depth increases. Within layer
$\ell$, $\mathbf{a}^{(\ell)}$ controls the combination described above: each
position contributes to the result in proportion to its current reliability, so
the received positions dominate early on and the missing ones are admitted only
gradually.

This schedule produces a coarse-to-fine completion. The early layers rely almost
entirely on the reliable received tokens and use them to fill in a first estimate
of the missing positions. Later layers, once those positions have been given
plausible values, gradually bring them into the reconstruction as well. In this
way the missing positions are recovered from trustworthy information first,
rather than from the empty placeholders, before being refined across the whole
grid. Finally, ADM adds the correction back to the
baseline $\mathbf{b}$, using separate strengths for the received and missing
positions selected by the mask $\mathbf{m}$,
\begin{equation}
  \widehat{\mathbf{z}}=\mathbf{b}
  +\alpha_{\mathrm{k}}\,\mathbf{m}\odot\Delta\mathbf{z}
  +\alpha_{\mathrm{u}}\,(1-\mathbf{m})\odot\Delta\mathbf{z},
  \label{eq:adm_gated_residual}
\end{equation}
where $\alpha_{\mathrm{k}},\alpha_{\mathrm{u}}$ are learnable positive scalars.
This lets the received codewords pass through almost untouched while the missing
positions, which must be inferred, receive a stronger correction.

\subsection{Prior-Conditioned Decoder}
\label{sec:decoder}
Once ADM has completed the latent, the decoder maps it back to the angle--delay
image,
\begin{equation}
  \widehat{\mathbf{X}}=D_{\phi}(\widehat{\mathbf{z}},\mathbf{P}),
  \label{eq:decode}
\end{equation}
following a U-Net expanding path. A $3\times3$ convolution first maps
$\widehat{\mathbf{z}}$ into the decoder's feature space, a bottleneck multi-head
self-attention block
\cite{transformer_vaswani2017} then mixes global context at the coarsest scale,
and finally $n_{\mathrm{down}}$ prior-conditioned upsampling blocks restore the
original resolution. The one departure from a standard U-Net concerns the skip
connections. A U-Net normally forwards the encoder's intermediate features to the
decoder through such skips, but here those features live at the UE and cannot
reach the BS without extra feedback overhead. We therefore replace the UE-side
skips with multi-scale features derived from the shared prior. A BS-side prior
encoder produces skip features
\begin{equation}
  \{\mathbf{p}^{\mathrm{skip}}_{\ell}\}_{\ell=1}^{n_{\mathrm{down}}}=E_P(\mathbf{P}),
  \label{eq:prior_skips}
\end{equation}
which carry high-resolution structure about the likely angle--delay support, while
a prior pyramid produces scale-matched SPADE maps
\begin{equation}
  \{\mathbf{p}^{\mathrm{mod}}_{\ell}\}_{\ell=1}^{n_{\mathrm{down}}}=\Pi_P(\mathbf{P}).
  \label{eq:prior_pyramid}
\end{equation}
Each upsampling block then combines these prior features with the current decoder
state,
\begin{equation}
  \mathbf{h}_{\ell+1}=\mathcal{U}_{\ell}\bigl(\mathbf{h}_{\ell},\mathbf{p}^{\mathrm{skip}}_{\ell},\mathbf{p}^{\mathrm{mod}}_{\ell}\bigr),
  \label{eq:decoder_upblock}
\end{equation}
where $\mathcal{U}_{\ell}$ upsamples $\mathbf{h}_{\ell}$, merges in the prior skip
feature, applies SPADE-modulated convolutions as in \eqref{eq:spade}, and adds a
residual shortcut. In this way the decoder is conditioned on the same prior as
the encoder and SAMM, yet reconstructs the image using only information already
available at the BS.

\subsection{Training Objective}
\label{sec:loss}
All modules are trained jointly and end-to-end under a single objective with four
terms: a reconstruction loss, a VQ commitment loss, an ADM latent-completion
loss, and a code-usage regularizer,
\begin{equation}
  \mathcal{L}=\lambda_{\mathrm{rec}}\mathcal{L}_{\mathrm{rec}}
  +\lambda_{\mathrm{cm}}\mathcal{L}_{\mathrm{cm}}
  +\lambda_{\mathrm{lat}}\mathcal{L}_{\mathrm{lat}}
  +\lambda_{\mathrm{cu}}\mathcal{L}_{\mathrm{cu}}.
  \label{eq:total_loss}
\end{equation}
The reconstruction loss $\mathcal{L}_{\mathrm{rec}}$ is the normalized MSE between the
reconstructed and ground-truth angle--delay images. The commitment loss
\begin{equation}
  \mathcal{L}_{\mathrm{cm}}=\frac{1}{|\mathcal{I}|}\sum_{i\in\mathcal{I}}
  \bigl\|\mathrm{sg}(\mathbf{z}_{q_i})-\widetilde{\mathbf{z}}_{{\rm e},i}\bigr\|_2^2
  \label{eq:commitment_loss}
\end{equation}
pulls each selected token toward its codeword, and where $\mathrm{sg}(\cdot)$ is the stop-gradient operator. The codewords themselves are
updated not by the usual codebook loss but by exponential moving averages of the
assigned encoder outputs \cite{oord2017neural}, which stabilizes the codebook
during training. The latent loss supervises ADM by comparing the recovered
latent $\widehat{\mathbf{z}}$ with the encoder latent $\mathbf{z}_{\rm e}$,
weighting the received and masked positions separately,
\begin{align}
  \mathcal{L}_{\mathrm{lat}}&=
  \omega_{\mathrm{r}}\,
  \frac{\bigl\|\mathbf{m}\odot\bigl(\widehat{\mathbf{z}}-\mathrm{sg}(\mathbf{z}_{\rm e})\bigr)\bigr\|_2^2}{\|\mathbf{m}\|_0}
  \nonumber \\
  &\quad \quad \quad \quad +\omega_{\mathrm{m}}\,
  \frac{\bigl\|(\mathbf{1}-\mathbf{m})\odot\bigl(\widehat{\mathbf{z}}-\mathrm{sg}(\mathbf{z}_{\rm e})\bigr)\bigr\|_2^2}{\|\mathbf{1}-\mathbf{m}\|_0}.
  \label{eq:latent_loss}
\end{align}
We set $\omega_{\mathrm{m}}>\omega_{\mathrm{r}}$, so the loss lets ADM keep the
received tokens intact while concentrating its effort on the harder missing
positions. The final term, the code-usage regularizer, prevents codebook collapse
by keeping the empirical codeword usage close to uniform
\cite{dieleman2018challenge,baevski2020wav2vec,zhang2023regularized}.
Let $\mathcal{B}$ be the set of selected tokens in a batch. The hard quantization
in \eqref{eq:vq} assigns each token to a single nearest codeword, which is not
differentiable and gives no signal about the other codewords. We therefore
replace it with a soft assignment
\begin{equation}
  \pi_{i,j}=\frac{\exp\!\bigl(-\|\widetilde{\mathbf{z}}_{{\rm e},i}-\mathbf{c}_{j}\|_2^2/\tau\bigr)}
  {\sum_{j'=1}^{J}\exp\!\bigl(-\|\widetilde{\mathbf{z}}_{{\rm e},i}-\mathbf{c}_{j'}\|_2^2/\tau\bigr)},
  \qquad
  \bar{p}_{j}=\frac{1}{|\mathcal{B}|}\sum_{i\in\mathcal{B}}\pi_{i,j},
  \label{eq:code_usage_prob}
\end{equation}
where $\pi_{i,j}$ is the (approximate) probability that token $i$ is quantized to
codeword $j$: it is largest for the nearest codeword and decays with distance,
controlled by the temperature $\tau=1$. Averaging $\pi_{i,j}$ over the batch
gives $\bar{p}_{j}$, the mean rate at which codeword $j$ is used. Stacking these
into $\bar{\mathbf{p}}=(\bar{p}_{1},\dots,\bar{p}_{J})$ yields the empirical usage
distribution over the codebook, and the penalty is its KL divergence from the
uniform distribution,
\begin{equation}
  \mathcal{L}_{\mathrm{cu}}=\sum_{j=1}^{J}\bar{p}_{j}\log\!\frac{\bar{p}_{j}}{1/J}
  =\log J-H(\bar{\mathbf{p}}),
  \label{eq:code_usage_loss}
\end{equation}
so that minimizing $\mathcal{L}_{\mathrm{cu}}$ maximizes the usage entropy and
keeps all codewords active.

\begin{algorithm}[t]
  \caption{Proposed PM-VQ CSI feedback scheme}
  \label{alg:proposed_scheme}
  \begin{algorithmic}[1]
    \Require CSI image $\mathbf{X}$, prior map $\mathbf{P}$, feedback dimension
    $\beta_{\mathrm{fb}}$
    \State Compute $k^{\star}$ from the budget constraint in \eqref{eq:k_star}
    \State Encode CSI and prior:
    $\mathbf{z}\gets E_{\theta}(\mathbf{X},\mathbf{P})$
    \State Generate latent-resolution prior $\widetilde{\mathbf{P}}$
    \State Score latent tokens:
    $\mathbf{s}\gets\mathrm{SAMM}(\mathbf{z},\widetilde{\mathbf{P}})$
    \State Select token positions:
    $\mathcal{I}\gets\operatorname*{top\text{-}k^{\star}}(\mathbf{s})$
    \For{$i\in\mathcal{I}$}
    \State Normalize and score-scale:
    $\widetilde{\mathbf{z}}_{i}\gets
    \mathrm{LN}(\mathbf{z}_{i}) s_i$
    \State Quantize:
    $q_i\gets\arg\min_j
    \|\widetilde{\mathbf{z}}_{i}-\mathbf{c}_j\|_2^2$
    \EndFor
    \State UE feeds back $\{q_i\}_{i\in\mathcal{I}}$ and $\mathcal{I}$
    \State BS reconstructs selected codewords:
    $\mathbf{z}_{q_i}\gets\mathbf{c}_{q_i}$
    \State Assemble sparse grid $\widetilde{\mathbf{z}}_0$ via \eqref{eq:adm_sparse_grid}
    \State Recover full latent:
    $\widehat{\mathbf{z}}\gets
    \mathrm{ADM}(\widetilde{\mathbf{z}}_0,{\mathbf{m}})$
    \State Decode CSI:
    $\widehat{\mathbf{X}}\gets D_{\phi}(\widehat{\mathbf{z}},\mathbf{P})$
    \State During training, compute $\mathcal{L}$ in \eqref{eq:total_loss} and update
    all trainable modules
  \end{algorithmic}
\end{algorithm}

\section{Dataset Generation}
\label{sec:dataset}
{The DL CSI dataset is generated with the Sionna link-level
simulator~\cite{hoydis2022sionna} under the 3GPP TR~38.901 urban
macrocell (UMa) non-line-of-sight (NLoS) channel
model~\cite{3gpp38901}. A single BS is located at the cell center,
$\mathbf{d}_{\rm BS}=[0,0,25]\ \mathrm{m}$, and each UE is placed at
\begin{equation}
    \mathbf{d}_{\rm UE}
    =
    [\,r\cos\theta,\;r\sin\theta,\;h_{\rm UE}\,]\ \mathrm{m},
    \label{eq:ue_position}
\end{equation}
where $r$ denotes the horizontal BS--UE distance, $\theta$ the azimuth
angle, and $h_{\rm UE}$ the UE height. These parameters are drawn
independently for each UE as
\begin{equation}
    \theta \sim \mathcal{U}[0,2\pi),\quad
    r \sim \mathcal{U}[35,250]\ \mathrm{m},\quad
    h_{\rm UE}\sim\mathcal{U}[1.5,22.5]\ \mathrm{m}.
    \label{eq:annulus_sampling}
\end{equation}}

The DL carrier frequency is $f_{\mathrm{DL}}=2.14\,\mathrm{GHz}$. The BS employs an $N_{\mathrm{BS}}=32$-element single-polarized uniform linear array (ULA) with the TR~38.901 antenna element pattern, while each UE is equipped with $N_{\mathrm{UE}}=4$ single-polarized antenna elements. The orthogonal frequency-division multiplexing (OFDM) grid contains $N_{\mathrm{RB}}=50$ resource blocks, each comprising $12$ subcarriers, with subcarrier spacing $\Delta f=30\,\mathrm{kHz}$. This yields $600$ subcarriers over an $18\,\mathrm{MHz}$ bandwidth. To emulate subband-based CSI reporting, only the center subcarrier of each resource block is retained, i.e., subcarrier $12r+6$ of resource block $r$ for $r=0,\ldots,N_{\mathrm{RB}}-1$. The resulting CSI representation therefore contains $N_{\mathrm{sc}}=50$ subcarriers with a spacing of $360\,\mathrm{kHz}$, sampled directly from the full-band channel response without interpolation.

For a given UE, the model first draws the large-scale parameters (path loss, shadow fading, delay spread, angular spreads, and Ricean $K$-factor) from the UMa distributions conditioned on the BS--UE topology, and then draws the cluster delays, cluster powers, and the per-cluster and per-ray departure/arrival angles. At each UE location we hold all of these quantities fixed, that is, the BS and UE positions, the large-scale parameters, and the entire cluster/ray geometry (delays, powers, and angles), and resample only the random initial phases of the rays across realizations. Each realization is therefore an \emph{independent phase realization conditioned on a fixed topology and a fixed cluster/ray geometry}, and the resulting variation is the coherent small-scale fading induced by the superposition of rays with independently drawn phases. The dataset is partitioned by UE location, with $7{,}500$ locations used for training and $750$ disjoint locations used for testing. For each location, $20$ independent phase realizations are generated, resulting in $150{,}000$ training samples and $15{,}000$ test samples. An additional phase-realization pool, disjoint from the training and test samples, is generated at every location solely for prior estimation.

\section{Baseline Methods}
\label{sec:baselines}

We compare the proposed scheme against three representative feedback baselines, all evaluated under the same uplink feedback dimension $\beta_{\rm fb}$ and UL SNR: a JSCC scheme, an SSCC scheme, and a training-free compressed-sensing scheme. The two learned baselines both adopt the transformer-based TransNet architecture \cite{cui2022transnet}, a widely used and competitive backbone for learning-based CSI feedback that captures long-range angle--delay dependencies through self-attention.

\subsection{Transformer-Based JSCC Baseline}
\label{sec:jscc}
The JSCC baseline adopts the TransNet-based feedback scheme \cite{cui2022transnet}. A transformer encoder maps the angle--delay channel $\mathbf{H}_{\rm ad}$ directly to $\beta_{\rm fb}$ complex symbols, normalized to unit average power per channel use, which are sent over the AWGN uplink,
\begin{align}
  \mathbf{y}_{\rm fb}=\mathbf{x}_{\rm fb}+\mathbf{n}_{\rm fb},
  \label{eq:jscc_channel}
\end{align}
where $\mathbf{x}_{\rm fb}\in\mathbb{C}^{\beta_{\rm fb}}$ are the transmitted symbols and $\mathbf{n}_{\rm fb}$ is complex AWGN whose variance is set by the UL SNR. A symmetric transformer decoder then reconstructs the channel from $\mathbf{y}_{\rm fb}$, and the encoder and decoder are trained jointly to minimize the channel NMSE. Because the encoder feeds analog symbols straight through the channel, a separate model must be trained for each $\beta_{\rm fb}$ and UL SNR.

\subsection{Transformer-Based SSCC Baseline}
\label{sec:sscc}

The SSCC baseline uses the same transformer encoder and decoder but, like the proposed method, feeds back a bit stream rather than analog symbols. The encoder produces a length-$\beta_{\rm out}$ latent, and each entry is passed through a sigmoid and a uniform $Q$-bit scalar quantizer, giving $\beta_{\rm out}Q$ bits in total. To match the reliable budget of \eqref{eq:budget} for $\beta_{\rm fb}$ channel uses at capacity $C_{\rm ul}$, the latent length is set to
\begin{equation}
  \beta_{\rm out}=\left\lfloor\frac{\beta_{\rm fb}C_{\rm ul}}{Q}\right\rfloor.
  \label{eq:sscc_bits}
\end{equation}
The model is trained end-to-end to minimize the channel NMSE. At the fixed budget $\beta_{\rm fb}C_{\rm ul}$, the number of quantization bits $Q$ trades resolution against dimensionality: a larger $Q$ quantizes each coefficient more finely but leaves room for fewer coefficients, whereas a smaller $Q$ keeps more coefficients at coarser resolution. We use $Q=2$.

\subsection{Orthogonal Matching Pursuit under Random Projection}
\label{sec:omp}

The final baseline is training-free and exploits the angle--delay sparsity of $\mathbf{H}_{\rm ad}$ directly. The channel is vectorized and split into real and imaginary parts to form $\mathbf{x}\in\mathbb{R}^{n}$, with $n=2N_{\rm sc}N_{\rm BS}N_{\rm UE}$. It is then compressed by a random sensing matrix $\mathbf{A}\in\mathbb{R}^{M\times n}$ with i.i.d.\ $\mathcal{N}(0,1/M)$ entries, where the number of real measurements $M=2\beta_{\rm fb}$ matches the feedback dimension. The measurements are received over the UL AWGN channel as
\begin{equation}
  \mathbf{y}=\mathbf{A}\mathbf{x} + \nv,
  \label{eq:cs_model}
\end{equation}
with noise variance set by the UL SNR. From $\mathbf{y}$, OMP greedily recovers $\mathbf{x}$ by solving $\min_{\mathbf{x}}\|\mathbf{y}-\mathbf{A}\mathbf{x}\|_{2}$ subject to a sparsity constraint $\|\mathbf{x}\|_{0}\le k_{\rm sp}$, with the sparsity level $k_{\rm sp}$ set from the empirical energy concentration of the channel. OMP thus serves as a sparsity-based reference that needs neither training nor channel statistics.

\section{Simulation Results}
\label{sec:results}
We evaluate the proposed PM-VQ scheme against the three baselines described in
Section~\ref{sec:baselines} on the UMa test dataset introduced in
Section~\ref{sec:dataset}. Unless otherwise stated, PM-VQ uses a test-time prior
power map estimated from $S=20$ independent channel realizations.
The prior-aided encoder maps each channel realization to $K=208$ latent tokens,
each of dimension $C=16$. These tokens are quantized using a learned codebook
with $J=512$ entries, so each transmitted codeword index requires
$\log_2 J=9$ bits. The codebook is updated by exponential moving average (EMA)
with a decay factor of $0.95$.
All PM-VQ components are trained jointly end-to-end for $500$ epochs with a
batch size of $125$, using the AdamW optimizer~\cite{loshchilov2017decoupled}
and a cosine learning-rate schedule. During training, the number of selected
tokens $k$ is sampled between $3$ and $200$ for each batch. This budget
randomization enables a single trained model to support multiple feedback
budgets by varying $k$ at inference. By contrast, the JSCC and SSCC neural
baselines are trained separately for each target combination of feedback dimension
and UL SNR using the same training dataset.
For a consistent comparison, all schemes are evaluated at the same UL SNR and
with the same feedback dimension of $\betafb$ complex UL channel uses.
Reconstruction accuracy is measured by NMSE.
The main architectural and training hyperparameters of PM-VQ are summarized in
Table~\ref{tab:hyperparams}.

To characterize performance across different scattering environments, we further split the test dataset according to the \emph{channel rank}. A high channel rank corresponds to a rich, \emph{diffuse} channel whose energy is spread across many angle--delay clusters, whereas a low rank corresponds to a \emph{sparse} channel dominated by one or two strong paths. We sort the test locations by their channel rank and partition them into four equal-sized quartiles $\mathrm{Q1}$--$\mathrm{Q4}$, ranging from the most diffuse ($\mathrm{Q1}$) to the most sparse ($\mathrm{Q4}$). The diffuse channels are the most challenging for CSI feedback, since their energy spreads across many non-negligible coefficients, precisely the regime that PM-VQ's prior-guided token selection and prior-conditioned reconstruction are designed to exploit. We therefore focus the simulation study on the $\mathrm{Q1}$ diffuse-channel dataset.

\begin{table}[t]
  \centering
  \caption{PM-VQ architecture and training hyperparameters.}
  \label{tab:hyperparams}
  \begin{tabular}{ll}
    \toprule
    \textbf{Hyperparameter} & \textbf{Value} \\
    \midrule
    \multicolumn{2}{l}{\textit{Architecture}} \\
    Latent tokens $K=hw$            & $208$ \\
    Latent channel dimension $C$    & $16$ \\
    Downsampling stages $n_{\mathrm{down}}$ & $3$ \\
    ADM self-attention blocks $L$   & $8$ \\
    Codebook size $J$               & $512$ \\
    \midrule
    \multicolumn{2}{l}{\textit{Training}} \\
    Optimizer                       & AdamW \\
    LR schedule                     & cosine ($\eta_{\min}=10^{-5}$) \\
    Peak LR (encoder / VQ)          & $1\times10^{-4}$ \\
    Peak LR (decoder / prior)       & $5\times10^{-5}$ \\
    Peak LR (ADM)                   & $5\times10^{-4}$ \\
    Peak LR (SAMM)                  & $2\times10^{-3}$ \\
    Weight decay                    & $1\times10^{-4}$ \\
    Gradient clip                   & $5.0$ \\
    Epochs                          & $500$ \\
    Batch size                      & $125$ \\
    Keep-count range $k$            & $[3,\,200]$ \\
    Codebook EMA decay              & $0.95$ \\
    Prior snapshots $S_{\mathrm{tr}}$ (training)  & $20$ \\
    \midrule
    \multicolumn{2}{l}{\textit{Loss weights}~\eqref{eq:total_loss},\,\eqref{eq:latent_loss}} \\
    $\lambda_{\mathrm{rec}}$        & $1.0$ \\
    $\lambda_{\mathrm{cm}}$   & $0.05$ \\
    $\lambda_{\mathrm{lat}}$        & $0.05$ \\
    $\lambda_{\mathrm{cu}}$         & $10^{-3}$ \\
    $\omega_{\mathrm{r}}$ / $\omega_{\mathrm{m}}$ & $0.1$ / $1.0$ \\
    \bottomrule
  \end{tabular}
  \vspace{-5mm}
\end{table}

\subsection{NMSE versus UL SNR at a Fixed Feedback Dimension}
\begin{figure*}[ht!]
  \centering
  \begin{subfigure}[b]{0.42\linewidth}
    \centering
    \includegraphics[width=\linewidth]{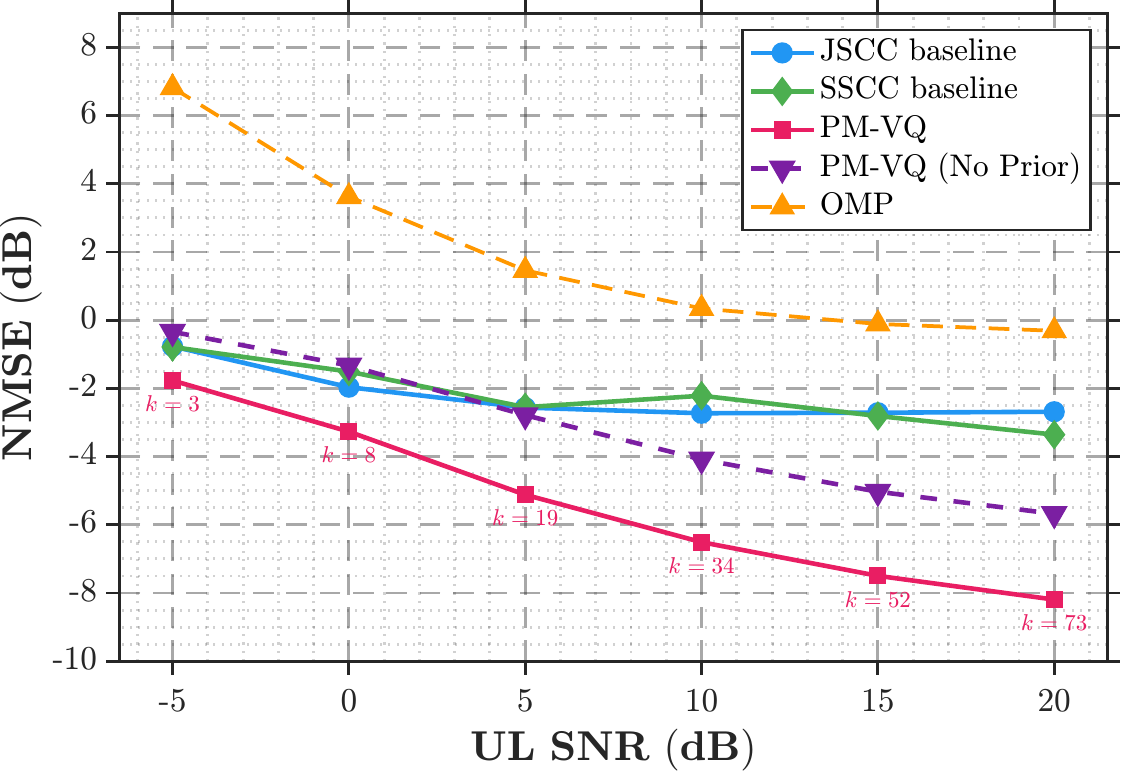}
    \caption{NMSE vs. UL SNR with $\beta_{\rm fb}=128$.}
    \label{fig:nmse_snr}
  \end{subfigure}
  \quad
  \begin{subfigure}[b]{0.5\linewidth}
    \centering
    \includegraphics[width=\linewidth]{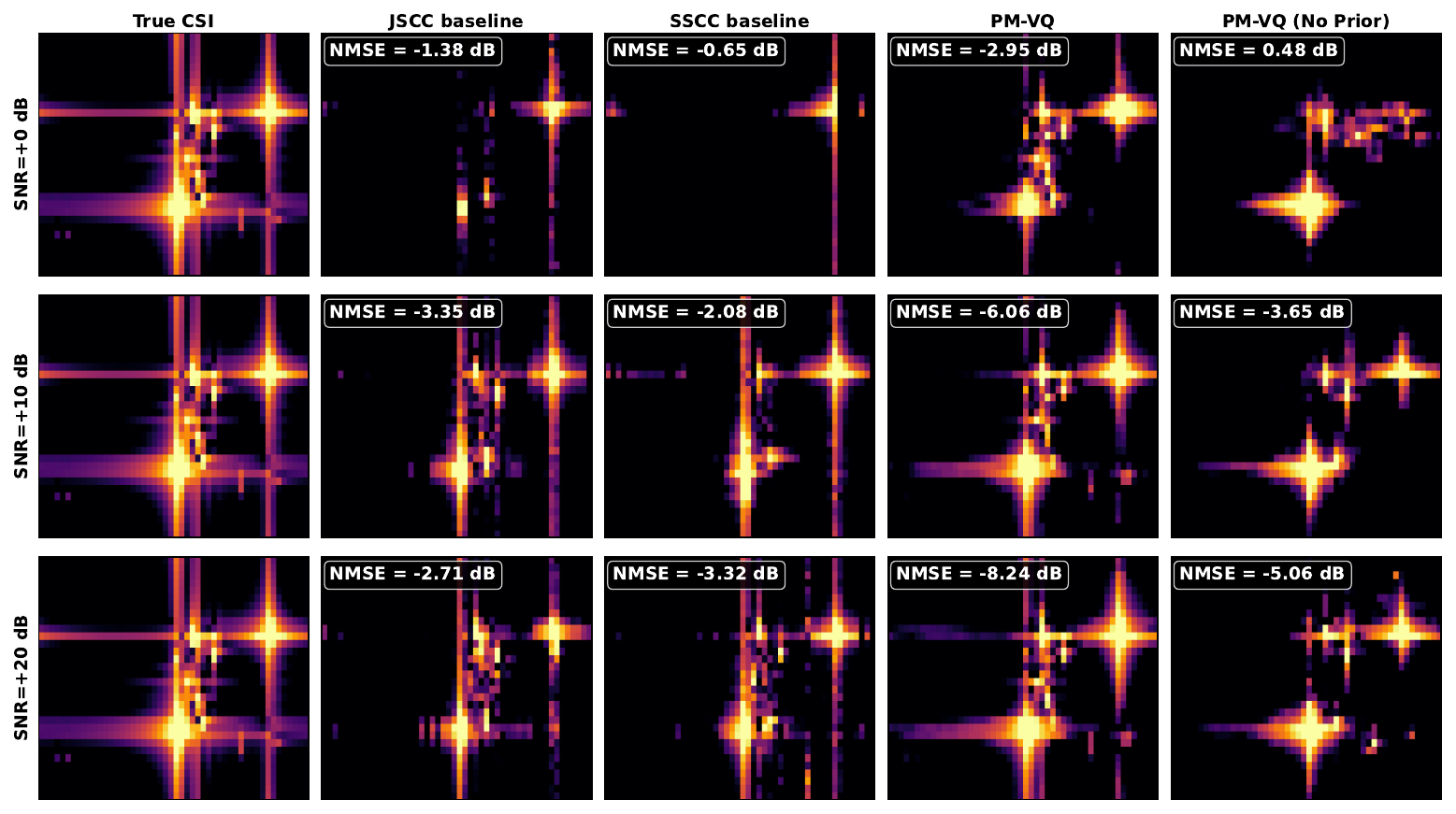}
    \caption{Angle--delay channel power heatmaps at
    UL SNR $\in\{0,10,20\}$~dB with $\beta_{\rm fb}=128$.}
    \label{fig:heatmap_snr}
  \end{subfigure}
  \caption{Comparison of CSI reconstruction across UL SNR.}
  \label{fig:eval_snr}
  \vspace{-5mm}
\end{figure*}

We first compare PM-VQ with the three baselines by sweeping the UL SNR from
$-5$ to $20$~dB at a fixed feedback
dimension $\beta_{\rm fb}=128$. To isolate the effect of prior conditioning, we
also evaluate PM-VQ (no prior), which uses the same architecture and selective
transmission mechanism but is trained and evaluated without prior information.
As shown in Fig.~\ref{fig:nmse_snr}, PM-VQ achieves the lowest NMSE at every
tested SNR, and its margin over the baselines increases as the UL channel
improves. The no-prior variant is less effective in the low-SNR regime but
progressively outperforms the learned baselines as the SNR increases. Among the
baselines, JSCC performs better over the low-to-moderate SNR range, whereas SSCC
becomes more competitive at high SNR. OMP has the highest NMSE throughout.

These trends reflect how the schemes use an increase in UL SNR. At fixed
$\beta_{\rm fb}$, JSCC always transmits the same number of analog symbols; a
higher SNR reduces the feedback noise but does not enlarge the transmitted
representation. Its curve therefore becomes nearly flat once the
fixed-dimensional representation, rather than the channel noise, becomes the
main bottleneck. SSCC instead converts the higher capacity into a larger
error-free bit budget. Because its scalar-quantization resolution is fixed, the
additional budget increases the number of transmitted coefficients, allowing
SSCC to continue improving after the JSCC curve begins to saturate. This
crossover reflects the different ways in which the two schemes exploit a better
UL channel, but it does not imply a general ordering because SSCC assumes
capacity-achieving channel coding and both learned baselines depend on their
architectures and training procedures.

PM-VQ shares this SSCC mechanism---a higher SNR buys a larger bit budget---but
couples it with content-adaptive token selection, latent completion, and prior
conditioning. At low SNR, only a small subset of latent tokens can be fed back,
so ADM must infer most of the latent representation from limited received
information. In this regime, the power-map prior is particularly valuable
because it supplies channel structure without consuming instantaneous feedback
bits. As the SNR increases, SAMM can forward more high-scoring tokens, enabling
the decoder to recover progressively finer channel details.

The comparison with PM-VQ (no prior) further separates the two sources of gain.
Under a very limited bit budget, selective transmission and latent completion
alone are insufficient to match the learned baselines, and the prior is
responsible for PM-VQ's clear advantage. Once the budget admits more tokens, the
no-prior variant itself surpasses the baselines, demonstrating the benefit of
content-adaptive token transmission and ADM. Full PM-VQ remains consistently
better, showing that prior-conditioned reconstruction continues to provide a
substantial improvement even when more latent information is available.

To complement the average NMSE curves, Fig.~\ref{fig:heatmap_snr} shows the
angle--delay channel power for a representative UE. The true channel contains
several clusters whose energy spreads over neighboring bins because the
continuous path angles and delays need not align with the discrete transform
grid. At low SNR, the baselines retain mainly the strongest components, whereas
full PM-VQ reconstructs more of the multi-cluster structure. Compared with
PM-VQ (no prior), its clusters are also aligned more accurately with the true
angle--delay locations, illustrating how the prior guides reconstruction when
only a few tokens can be transmitted. As the SNR increases, SSCC and both PM-VQ
variants recover additional structure, while JSCC remains limited by its fixed
representation. Full PM-VQ preserves the weak components most faithfully and
introduces fewer artifacts, consistent with its leading NMSE curve.

\subsection{NMSE versus Feedback Dimension at a Fixed UL SNR}
\begin{figure*}[ht!]
  \centering
  \begin{subfigure}[b]{0.42\linewidth}
    \centering
    \includegraphics[width=\linewidth]{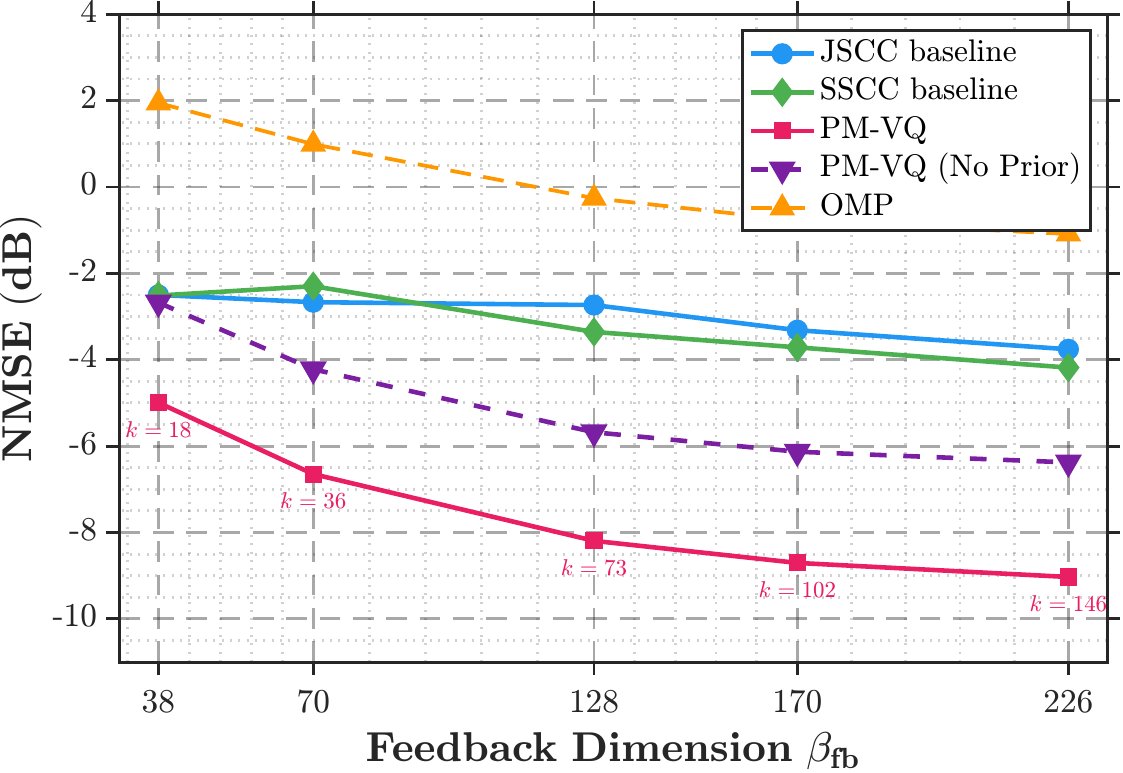}
    \caption{NMSE vs. $\beta_{\rm fb}$ at UL SNR $=20$~dB.}
    \label{fig:nmse_fb}
  \end{subfigure}
  \quad
  \begin{subfigure}[b]{0.5\linewidth}
    \centering
    \includegraphics[width=\linewidth]{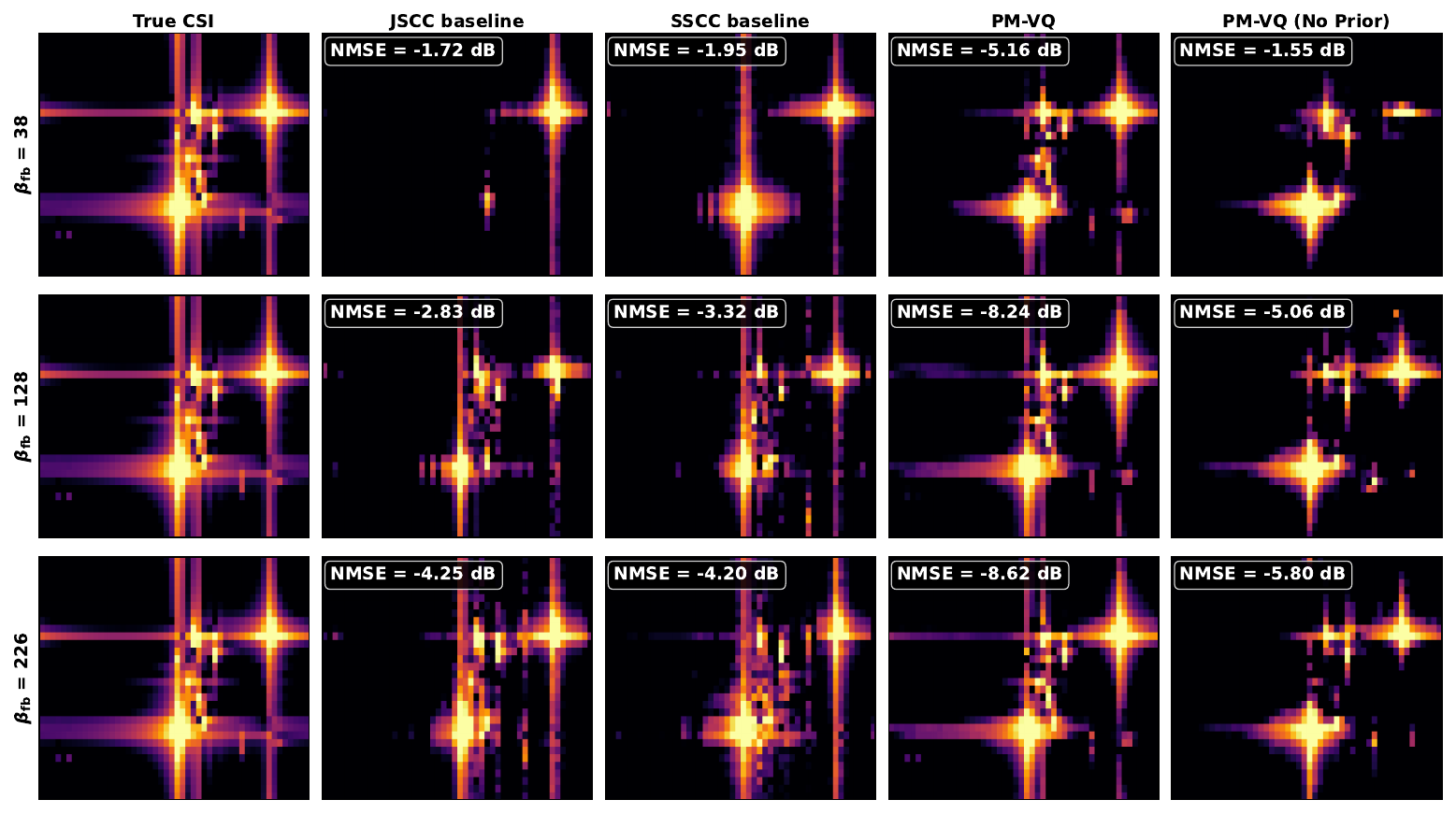}
    \caption{Angle--delay channel power heatmaps for
    $\beta_{\rm fb}\in\{38,128,226\}$ at UL SNR $=20$~dB.}
    \label{fig:heatmap_fb}
  \end{subfigure}
  \caption{Comparison of CSI reconstruction across feedback dimension.}
  \label{fig:eval_fb}

    \vspace{-5mm}
      \end{figure*}

Next, we fix the UL SNR at $20$~dB and vary the feedback dimension
$\beta_{\rm fb}$, defined as the number of complex UL channel uses. As shown in
Fig.~\ref{fig:nmse_fb}, the NMSE generally decreases as $\beta_{\rm fb}$
increases. Full PM-VQ achieves the lowest NMSE at every feedback
dimension, followed by PM-VQ (no prior), and their advantages over the baselines
become more pronounced as $\beta_{\rm fb}$ increases. Among the learned
baselines, JSCC and SSCC have
similar performance under the smallest budget. JSCC initially improves more,
whereas SSCC benefits more strongly once the available feedback becomes larger.
OMP improves with $\beta_{\rm fb}$ but remains the worst-performing method
throughout.

At the high UL SNR of $20$~dB, changing $\beta_{\rm fb}$ primarily changes how
much channel information each scheme can convey. For JSCC, a larger feedback
dimension allows more UL modulated symbols to be transmitted. For SSCC, the
reliable bit budget grows proportionally, allowing more quantized coefficients
to reach the BS. For OMP, the larger dimension provides more random
measurements. Although the diffuse channels remain sparse in the angle--delay
domain, their energy is distributed over more significant coefficients, and
off-grid leakage spreads each physical path across neighboring bins. This higher
effective sparsity makes greedy support recovery more difficult and limits the
benefit of the additional measurements. In PM-VQ, a larger feedback dimension
allows SAMM to transmit more selected tokens. The initial gains are substantial,
but they diminish once the most informative tokens have already been conveyed.

We next compare PM-VQ with PM-VQ (no prior) to distinguish the contributions of
selective latent-token transmission and prior conditioning. Under the tightest
feedback dimension, the no-prior variant performs comparably to the learned
baselines. Its advantage grows as the feedback dimension increases, showing that
selective token transmission and latent completion become increasingly effective
when more informative tokens can be conveyed. The consistent gap between full
PM-VQ and its no-prior variant reveals the additional benefit of the power-map
prior across the entire feedback range. Full PM-VQ therefore remains the
best-performing scheme throughout the sweep.

Finally, Fig.~\ref{fig:heatmap_fb} qualitatively compares the reconstructions of
the same representative channel used in Fig.~\ref{fig:heatmap_snr}. At
the smallest feedback dimension, JSCC recovers mainly one dominant cluster,
whereas SSCC recovers the two dominant clusters but misses much of the weaker
structure. PM-VQ (no prior) also misplaces or merges some components. Full
PM-VQ uses the prior to localize the dominant clusters closer to their true
angle--delay positions and preserve more of the weak structure. As the feedback
dimension increases, all schemes recover more channel detail, but full PM-VQ
maintains the most accurate cluster locations and the least spurious energy.
The comparatively small visual change at the largest dimensions is consistent
with the flattening of its average NMSE curve. Overall, the qualitative
reconstructions follow the same performance ordering as the NMSE results in
Fig.~\ref{fig:nmse_fb}.

\subsection{Ablation Study of Prior Side Information}

\begin{figure*}[t]
  \centering
  \begin{subfigure}[t]{0.32\textwidth}
    \centering
    \includegraphics[width=\linewidth]{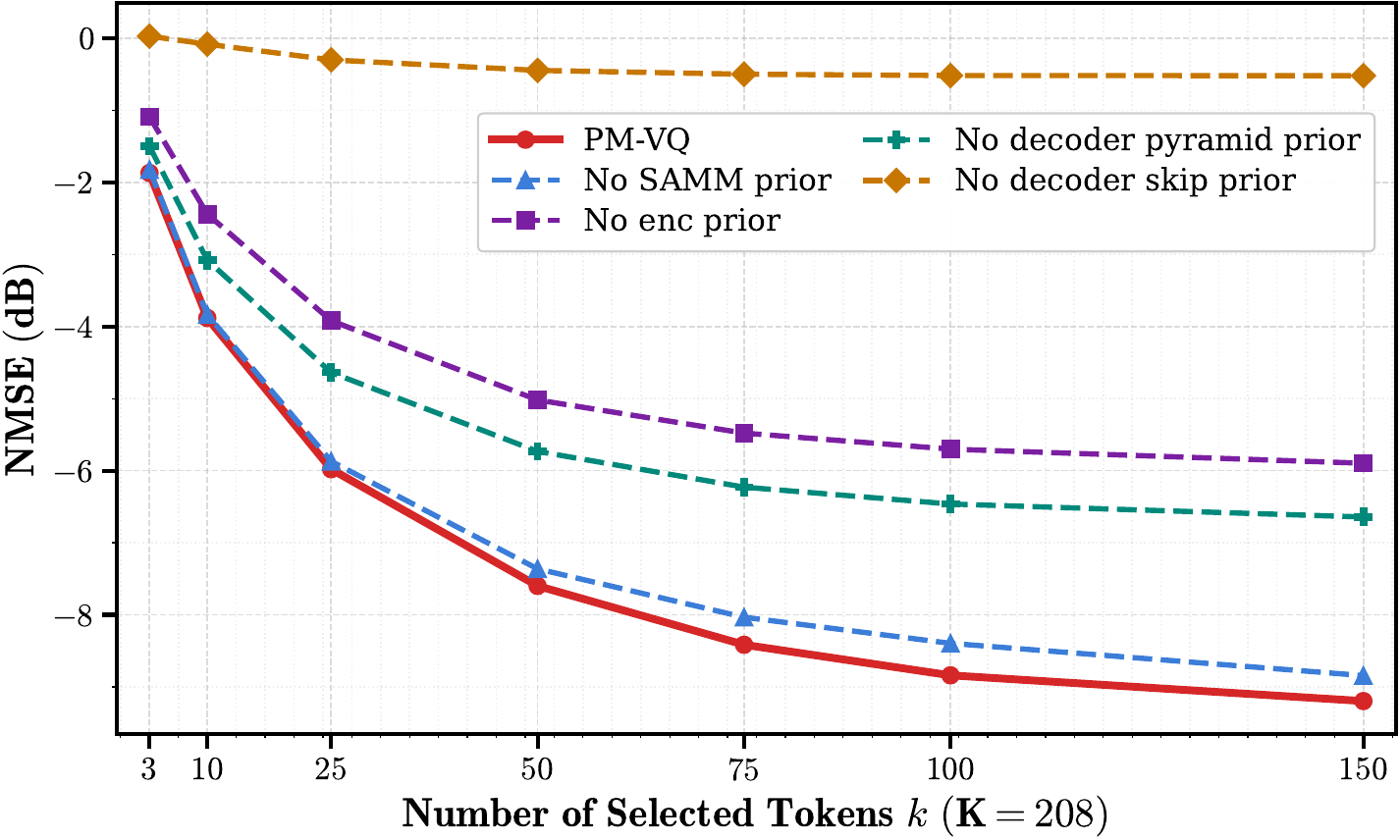}
    \caption{Effect of removing each prior-conditioning pathway at inference.}
    \label{fig:prior_ablation}
  \end{subfigure}
  \hfill
  \begin{subfigure}[t]{0.32\textwidth}
    \centering
    \includegraphics[width=\linewidth]{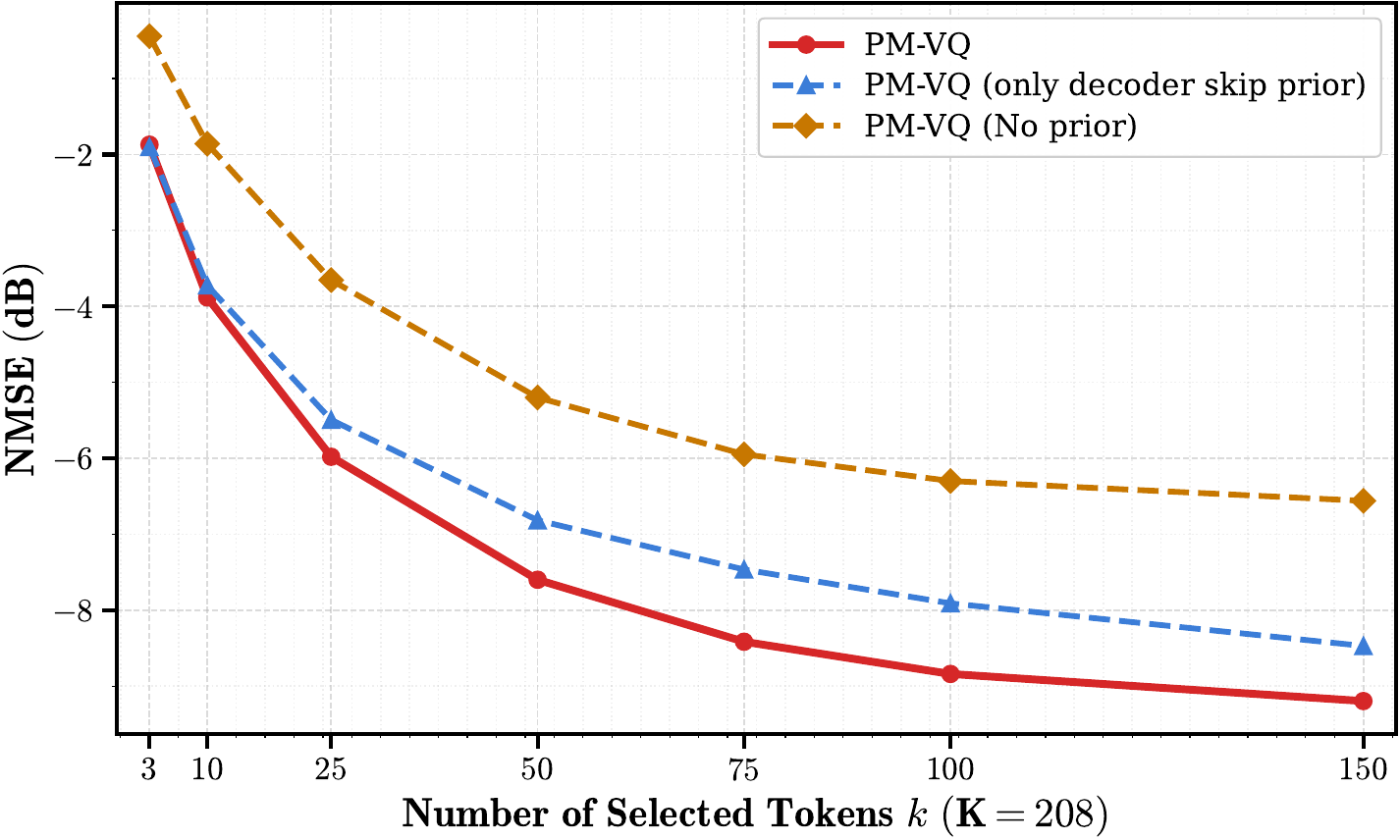}
    \caption{Comparison of models trained with different prior configurations.}
    \label{fig:prior_ablation_2}
  \end{subfigure}
  \hfill
  \begin{subfigure}[t]{0.32\textwidth}
    \centering
    \includegraphics[width=\linewidth]{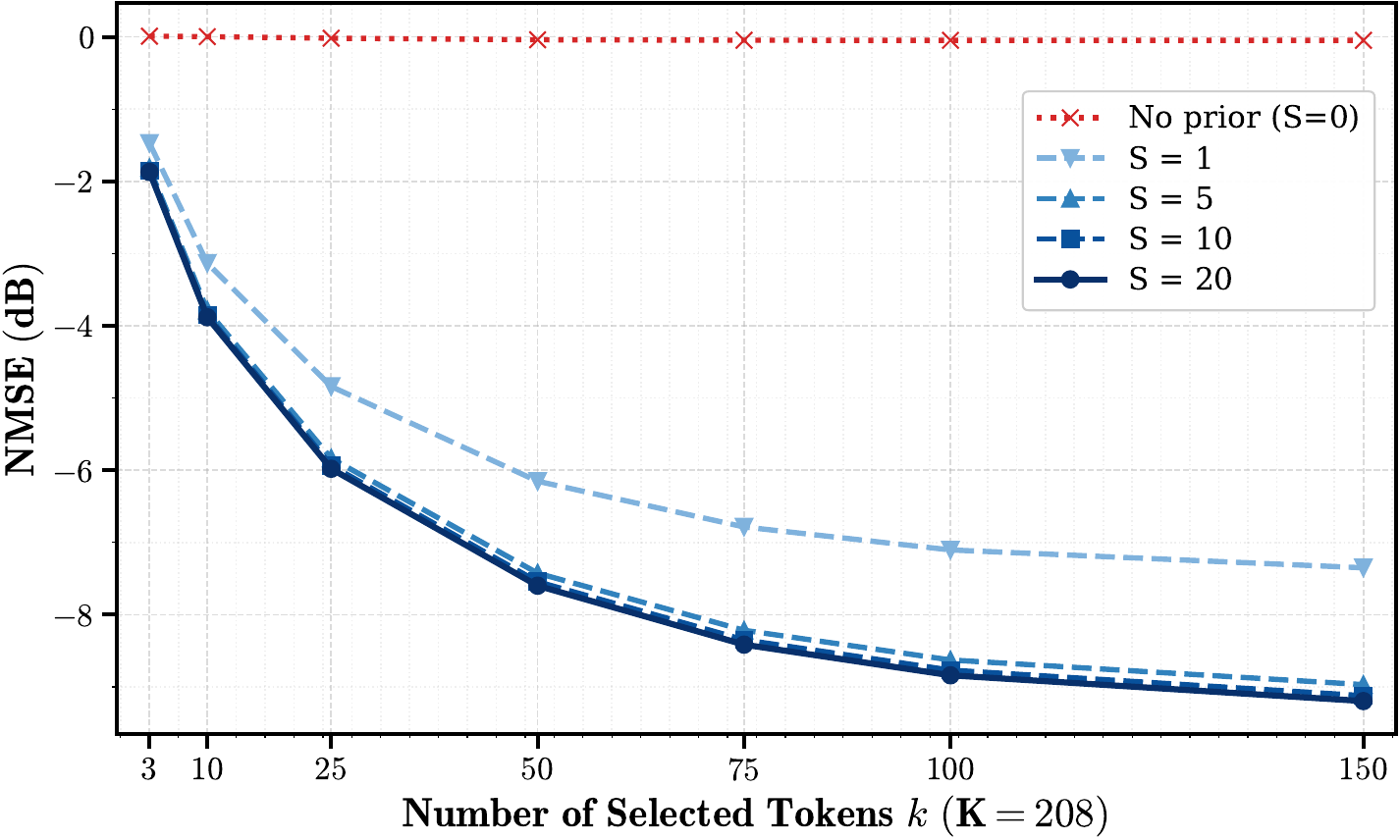}
    \caption{Effect of prior-estimation quality.}
    \label{fig:sample_number}
  \end{subfigure}
  \caption{Ablation of prior side information in PM-VQ.}
  \label{fig:prior}
    \vspace{-5mm}
\end{figure*}

The preceding NMSE studies establish that prior side information and
content-adaptive token selection both improve CSI reconstruction. To understand
where the prior gain comes from, Fig.~\ref{fig:prior} examines the prior from three
complementary perspectives. Figure~\ref{fig:prior_ablation} measures how strongly
a single trained model relies on each prior-conditioning pathway when that pathway
is disabled at inference, Fig.~\ref{fig:prior_ablation_2} retrains the model from
scratch under different prior configurations to confirm these conclusions hold
beyond a single checkpoint, and Fig.~\ref{fig:sample_number} examines how
accurately the prior must be estimated at inference.

For the module-wise ablation, one pathway is removed at inference while all
network weights remain fixed. The resulting performance loss therefore reflects
how strongly the trained model depends on that pathway. The decoder-skip pathway
is the most important: removing it eliminates most of the reconstruction gain
across the feedback range. Encoder conditioning has the next largest effect,
while the decoder-pyramid pathway provides a smaller but consistent
improvement. In contrast, removing direct prior conditioning from SAMM changes
the performance very little. This suggests that the prior-aided encoder has
already embedded sufficient statistical information in the latent tokens for
SAMM to rank them effectively. ADM is not included because it is prior-free by
design and completes the latent solely from the received tokens and their
positions.

This ordering follows from the asymmetric U-Net architecture. The UE-side
encoder features cannot be sent to the BS without additional feedback, so the
prior-derived skip features are the decoder's only source of high-resolution,
multi-scale side information. Without them, the decoder must recover fine
channel structure from the low-resolution latent alone. The encoder prior shapes
the latent before selection and quantization, while the decoder-pyramid prior
refines the features during upsampling. Thus, the trained model uses the prior
mainly for reconstruction at the decoder, especially through the prior-derived
skip connections.

Next, we verify whether these conclusions still hold when the network is
retrained, since the above test only disables paths in a single trained model.
In Fig.~\ref{fig:prior_ablation_2}, we compare three models trained from scratch:
the full PM-VQ model, a model that uses only the decoder-skip prior, and a model
that uses no prior. The ordering is consistent with the
inference-time result: the full model achieves the lowest NMSE for all $k$, the
no-prior model is the worst, and the decoder-skip-only model stays much closer to
the full model than to the no-prior one, recovering most of the full-prior gain.
This confirms that the decoder-skip path carries most of the benefit of the
prior, while the remaining paths provide a smaller but consistent improvement.

Having identified the decoder-skip pathway as the main source of prior gain, we
next ask how accurately the prior must be estimated. Withholding it at inference
causes reconstruction to collapse, consistent with the strong dependence on
prior-derived decoder features. This test reflects how a prior-trained model
responds to missing side information and is not equivalent to training a model
without the prior. Nevertheless, even a rough power-map estimate restores most
of the lost performance, and modest averaging approaches the result obtained
with a well-estimated prior. This indicates that PM-VQ mainly needs the coarse
angle--delay support; more accurate averaging primarily refines the power
values. The model is therefore robust to moderate prior-estimation error but
sensitive to the complete absence of the prior.

\subsection{Ablation Study of the Token Selection Mechanism}

\begin{figure}[t]
  \centering
  \includegraphics[width=0.95\linewidth]{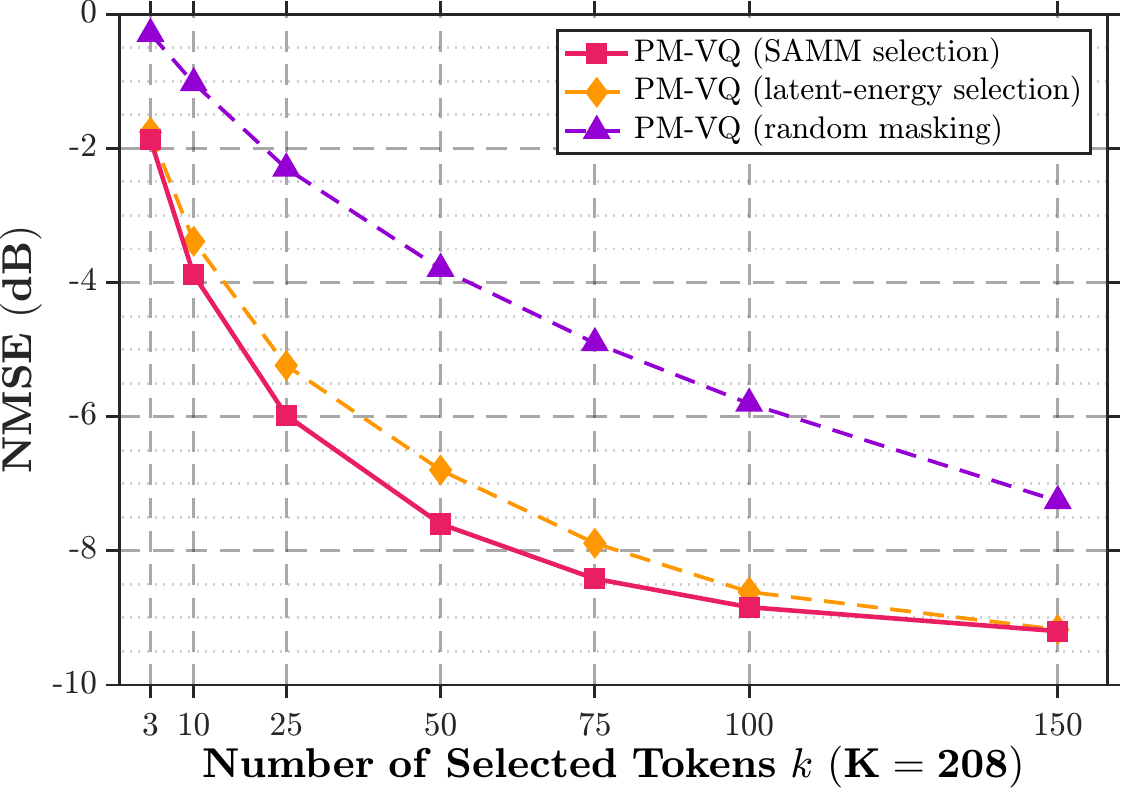}
  \caption{Effect of the token-selection rule in PM-VQ.}
  \label{fig:random_mask}
  \vspace{-5mm}
\end{figure}

To study the effect of the token selection mechanism, Fig.~\ref{fig:random_mask} compares
the learned SAMM rule with two alternatives while keeping the remaining PM-VQ
pipeline unchanged. Latent-energy selection retains the tokens with the largest
magnitudes, whereas random masking selects the transmitted positions uniformly
at random.
SAMM achieves the best performance across the feedback range. Latent-energy
selection is a strong heuristic and approaches SAMM when the feedback dimension is either
very tight or very large, but it falls behind most clearly at intermediate
feedback dimensions. Random masking performs substantially worse throughout, confirming that
the identities of the transmitted tokens, rather than only their number, are
important for reconstruction.

This behavior follows from the information available at each feedback dimension. Under a
very tight constraint, both SAMM and the energy rule tend to select the few
dominant tokens. With a large feedback dimension, the most informative tokens can be retained,
so the two rules again become similar. At intermediate feedback dimensions, however, the
selector must balance token strength, spatial coverage, and redundancy with the
tokens that ADM can reliably complete. Magnitude alone does not capture these
interactions, whereas SAMM is trained jointly with ADM and the decoder to score
tokens according to their reconstruction value. Learned content-adaptive
selection therefore provides its clearest benefit in this intermediate regime.

\section{Conclusion}
\label{sec:conclusion}
We proposed PM-VQ, a variable-rate SSCC scheme for FDD CSI feedback that
uses the average angle--delay power map as side information. SAMM selects the
latent tokens to be vector-quantized and transmitted, while the prior-free ADM
completes the missing latent positions at the BS. Budget-randomized training
allows the same model to operate over multiple feedback dimensions without
retraining.
PM-VQ outperforms the baselines across all tested UL SNRs and feedback
dimensions in the challenging diffuse regime.
The ablations show that the prior gain comes mainly from the decoder-skip
pathway; encoder and decoder-pyramid conditioning provide secondary gains,
whereas direct prior input to SAMM adds little. A coarse power-map
estimate is sufficient for most of the benefit, but removing the prior entirely
causes a performance loss. In contrast, SAMM's learned selection rule remains
important and outperforms both latent-energy selection and random masking,
especially at intermediate feedback dimensions. These findings suggest a clear
design principle: use the prior primarily to guide reconstruction at the BS and
use learned content-adaptive selection to allocate the instantaneous feedback
budget.

{\small
  \bibliographystyle{IEEEtran}
  \bibliography{references}
}

\end{document}